\documentclass[reprint,tightenlines,aps, pra, floats,nofootinbib,amssymb,groupedaddress]{revtex4-1}

\usepackage[sort&compress]{natbib}
\usepackage{epsf,epsfig, bm}
\usepackage{amsmath}
\usepackage{amsfonts,graphicx,soul,mathrsfs}
\usepackage{hyperref, comment}
\usepackage{subfigure}
\usepackage[normalem]{ulem}

\usepackage[dvipsnames]{xcolor}
\usepackage{color,colortbl,mathrsfs,pgf,tikz}
\usepackage[compat=1.1.0]{tikz-feynman}

\newcommand{\be}{\begin{equation}}
 \newcommand{\ee}{\end{equation}}
\newcommand{\bear}{\be\begin{array}}
\newcommand{\bea}{\begin{eqnarray}}
\newcommand{\eea}{\end{eqnarray}}
\newcommand{\nn}{\nonumber}

\newcommand{\bee}{{\bf e}}
\newcommand{\bp}{{\bf p}}
\newcommand{\br}{{\bf r}}

\newcommand{\bk}{{\bf k}}

\newcommand{\la}{\langle}
\newcommand{\ra}{\rangle}

\usepackage[dvipsnames]{xcolor}

\newcommand{\dst}{\displaystyle}
\newcommand{\fr}[2]{\frac{{\dst #1}}{{\dst #2}}}

\begin{document}
\title{
Vavilov-Cherenkov emission with a twist: \\ a study of the final entangled state
}
\author{A.D. Chaikovskaia}
\email{adkatanaeva@itmo.ru}
\affiliation{School of Physics and Engineering, ITMO University, 197101 St.~Petersburg, Russia}
\author{D.V. Karlovets}
\affiliation{School of Physics and Engineering, ITMO University, 197101 St.~Petersburg, Russia}
\author{V.G.~Serbo}
\affiliation{Novosibirsk State University, 630090 Novosibirsk, Russia\\
Sobolev Institute of Mathematics, 630090 Novosibirsk, Russia}
\affiliation{School of Physics and Engineering, ITMO University, 197101 St.~Petersburg, Russia}

\thispagestyle{empty}

\date{\today}
\vspace*{3cm}

\begin{abstract}
{ We present a theoretical investigation of the Vavilov-Cherenkov (VC) radiation by a  plane-wave or  twisted electron. Special emphasis is put on the question whether and at what conditions the emitted VC photons can be twisted. For this aim we obtain a general expression in the coordinate and momentum representations for the quantum state of the final electron-photon system that is a result of the radiation process itself and does not depend on the properties of a detector. It is shown that this {\it evolved state} is an entangled state of an electron and a photon, and both particles can be twisted.  
A direct consequence of this result follows: if one uses a detector  sensitive to the twisted electron (photon) with the definite projection of the total angular momentum (TAM), then the final photon (electron) also will be in the twisted state with a definite TAM projection. Further, we investigate 
the polarization properties of the final twisted photon in more general conditions than has been calculated before. Finally, we exploit a close similarity between the discussed VC radiation and the process of the equivalent photon emission in the Weizs\"acker-Williams method and find the corresponding final state.
}
\end{abstract}
\maketitle
\section{Introduction}

{The Vavilov--Cherenkov (VC) radiation was discovered in 1934~\cite{discovery} and very soon explained by Frank and Tamm~\cite{Tamm-Frank} in the framework of classical electrodynamics.
A few years later, Ginzburg~\cite{VLG1940} and Sokolov~\cite{sokolov1940}  gave the quantum derivation of this phenomenon and found quantum corrections to the classical
Frank--Tamm result.}

Historical and contemporary reviews on the subject can be found in \cite{Bolotovsky-2009, VLG, Afanasyev}.
A renewed motivation for the study of VC radiation within quantum electrodynamics (QED) is brought by recent theoretical and experimental advances with the so-called {\it twisted} particles, i.e. those carrying nonzero orbital angular momentum (OAM). The reviews~\cite{Blioch-Ivanov-2017, Knyazev-Serbo-2018, Ivanov:2022jzh} give an ample discussion on the properties of twisted electrons and photons and their experimental status.
A QED description of VC radiation emitted by a twisted electron in the transparent medium has recently been considered in~\cite{Kaminer-2016, ISZ-2016}. There, the analytical expressions have been obtained for the spectral and spectral-angular distributions as well as for the polarization properties of the emitted radiation. 


In any quantum process the detection scheme plays a crucial role in the measurement of the final quantum state characteristics. In the previous studies of VC radiation the most common method is to perform calculations for the final states being detected in the form of plane waves both for the electron and the photon as in Refs.~\cite{VLG1940, sokolov1940, Kaminer-2016, ISZ-2016}.  Another approach~\cite{KBGSS-2022, KBGSS-2023}, based on the so-called generalized measurements, focuses on the entanglement between a pair of the final particles and implies in practice the usage of a certain post-selection protocol for the final electron only when the momentum of the electron is measured with a large uncertainty. In this approach, the {\it evolved} state of the emitted photon appears.

In the present work we aim to give a complementary approach that allows one to perform a \textit{quantum tomography} of the scattered radiation without the need to introduce detector states.
In particular, we develop an exact analytical expression for the final system wave function in the coordinate and momentum representations that is {\it irrespective of the detector properties}. We use the well--established $S$--matrix theory of the evolved state that provides the wave function of the outgoing photon as it results from the scattering process itself.  

{ Describing the final state with an explicit wave function can be quite useful for solving a number of problems. 
The corresponding wave function of the final system in Ref.~\cite{ISZh-2019} was used to solve some important questions related to the application of the Landau-Yang theorem for twisted photons. The wave function of the final state has also been recently used in Ref.~\cite{KSS-2021} for the description of the resonance scattering of twisted laser photons on the ultra-relativistic partially striped ions, which is a basic process in the project of the Gamma Factory at the Large Hadron Collider.}

As a part of motivation for this study a few words on the somewhat natural relation between twisted photons and VC radiation are in order.
The VC radiation has a very specific angular distribution of final photons of energy $ \omega $: they are concentrated in a narrow cone near the polar angle
 \be
\theta_\gamma \approx \arccos \left (\fr{1}{vn(\omega)} \right),
\label{1}
 \ee
where $ v $ is the velocity of the emitting electron moving along the $ z $-axis
and { $n(\omega)$ is the refraction index of the transparent media}. Just the same feature is possessed by the wave function of a twisted photon, for which an {\it opening (or conical) angle} is equal to
 \be
\theta_\gamma= \arccos \left (\fr{k_z}{| \bk |} \right) = \arccos \left (\fr{1}{vn(\omega)} \right),
 \ee
meaning that it is determined by the magnitude of the longitudinal momentum of the photon $ k_z = \omega / v $ and its energy $ {\omega = | \bk | / n(\omega)} $. 
Second, VC radiation is linearly polarized in the scattering plane. As we show below, our result for the case of the low energy photons reproduces this observation.

The paper is organized as follows. We start with some general formalism: introducing  the notion of the evolved wave function of the final system, the basic notations and the standard expression for the amplitude of the VC process in Section~\ref{Sec_Evolved}.
Then, in Section~\ref{Sec_PW} we perform an analysis for the case of an ordinary plane-wave initial electron and derive the evolved wave function of the quantum system in the coordinate and momentum representations that corresponds to the entangled state of an emitted photon and an outgoing electron. It is remarkable that a symmetrical and relatively simple  wave function could be obtained.
In Section~\ref{Sec_TW} we generalize the calculation for the case of the initial twisted electron represented as a proper superposition of the plane waves comprising a Bessel wave state.  { In Section~\ref{Sec_WW}  we
develop 
a parallel to the issue of a virtual photon emission in the deep inelastic $ep$-scattering and to the emission of an equivalent photon in the Weizs\"acker-Williams method (see Refs. \cite{BLP} and \cite{BGMS-1975}). }
We discuss the impact of the achieved result and its possible implications in Section~\ref{Sec_Disc}.  { In conclusion we emphasize some points discussed in the paper.}  Through the paper we use the relativistic system of units
$ \hbar = 1 $, $ c = 1 $ and $ \alpha=e^2/(\hbar c)\approx 1/137 $, where $e$ is the electron charge.

\section{General formalism}
\label{Sec_Evolved} 
\subsection{Definition of evolved state}
\label{Sec_Evolved_def} 
Let us start with emission of a photon by a charged particle (for definiteness, an electron), 
\be
e \to e' + \gamma,
\ee 
in the lowest order of the perturbation theory in QED. It can generate either VC radiation or transition radiation in medium, synchrotron radiation, undulator radiation in a given electromagnetic field, and so on. A quantum state of the final system as it evolves from the reaction irrespectively of the measurement protocol can be called \textit{pre-selected} or \textit{evolved} because no measurements have been done yet. An initial state $|\text{in}\ra$ of the electron and an evolved two-particle state of the final system of the electron and the photon are connected via an evolution operator $\hat{S}$ \cite{BLP}
\be
\hat{S} = \text{T} \exp\left\{-ie\int d^4x\, \hat{j}_{\mu} \hat{A}^{\mu}\right\}
\label{Sev}
\ee
as follows:
\be
|e', \gamma\ra^{(\text{ev})} = \hat{S}\,|\text{in}\ra = \sum\limits_{f} |f\ra \la f|\hat{S}|\text{in}\ra,
\label{2part}
\ee
where we have expanded the unitary operator over a complete set of two-particle states (with no virtual particles on the tree-level)
\bea
\hat{1} = \sum\limits_{f} |f\ra \la f| = \sum\limits_{f_e}\sum\limits_{f_{\gamma}}|f_e, f_{\gamma}\ra \la f_e, f_{\gamma}|.
\eea
In particular, this can be the plane-wave states with the definite four-momenta 
$p'=(E',{\bf p}'),\ k=(\omega, {\bf k})$ and the helicities $\lambda'= \pm 1/2, \lambda_{\gamma} = \pm 1$ for $e'$ and $\gamma$, respectively, so that
\be
\hat{1} = \sum\limits_{\lambda' \lambda_{\gamma}} \int \frac{d^3p'}{(2\pi)^3} \frac{d^3k}{(2\pi)^3} |\bp'\lambda',\bk\lambda_{\gamma} \ra\la\bp'\lambda',\bk\lambda_{\gamma}|.
\label{ComplPW}
\ee 
In this case, the average
\be
S^{(1)}_{fi} = \la f_e, f_{\gamma}|\hat{S}|\text{in}\ra = \la\bp'\lambda',\bk\lambda_{\gamma}|\hat{S}|\text{in}\ra
\label{SPW}
\ee 
is a customary first-order matrix element with two final plane-wave states
whose polarization properties are described by bispinor $u_{\bp'\lambda'}$ 
and four-vector $e_{\bk \lambda_\gamma}=(0,{\bf e}_{{\bf k} \lambda_\gamma})$.

Let us first assume that no final particle is detected and discuss how one can define ``the wave function'' of the entangled evolved state. The field operators in the Heisenberg representation of the photon  are expanded into series of the creation and annihilation operators of the plane-wave states
\bea
 \hat{\bf A}(x) = \sum\limits_{\lambda_{\gamma}=\pm 1} \int\frac{d^3k}{(2\pi)^3} \left({\bf A}_{{\bf k}\lambda_{\gamma}}(x)\, \hat{c}_{{\bf k}\lambda_{\gamma}} + \text{h.c.}\right),\\
 {\bf A}_{{\bf k}\lambda_{\gamma}}(x) = \frac{\sqrt{4\pi}\,{\bf e}_{{\bf k}\lambda_{\gamma}} }{\sqrt{2\omega}\,n}\, e^{-i kx},
  \label{Fopg}
\eea
where $x=(t, \br)$ and  $n=n(\omega)$ is the refraction index of medium. Analogously, for the electron we have 
 \bea
\hat\psi(x) = \sum\limits_{\lambda'=\pm 1/2} \int\frac{d^3p'}{(2\pi)^3} \psi_{{\bf p}'\lambda'}(x) \hat{a}_{{\bf p}'\lambda'},
\\ \psi_{{\bf p}'\lambda'}(x) = \fr{u_{{\bf p}'\lambda'}}{\sqrt{2 E'}}\,e^{-ip'x},
 \label{FOpe}
\eea
where we have omitted the positron part.
Now, following the standard interpretation \cite{BLP}, the two-particle evolved state in four-dimensional space-time looks as follows (recall Eq.~(\ref{2part})):
\bea
\label{Psi-coor-1}
&& \la 0| \hat{\psi}(x_e) \hat{{\bf A}}(x_{\gamma})|e',\gamma\ra^{(\text{ev})}   
 \\\
 && =\int\frac{d^3p'}{(2\pi)^3}\frac{d^3k}{(2\pi)^3}\,\sum\limits_{\lambda' \lambda_{\gamma}}
\fr{u_{{\bf p}'\lambda'}}{\sqrt{2 E'}}\,e^{-ip'x_e}\,
\nn
\\\
 && \times \fr{\sqrt{4\pi}\,{\bf e}_{{\bf k}\lambda_{\gamma}} }{\sqrt{2\omega} n}\, e^{-i kx_\gamma}  S_{fi}^{(1)},
 \nn
 \eea
with $S_{fi}^{(1)}$ constructed following the Feynman rules, see Eq.~(\ref{SPW}) further. So the combination 
\bea
\sum\limits_{\lambda' \lambda_{\gamma}}  u_{{\bf p}'\lambda'}\, {\bf e}_{{\bf k}\lambda_{\gamma}}\, S_{fi}^{(1)}
\label{APsievk}
\eea
can be called the wave function of the two-particle evolved state in the momentum representation. Clearly, it bears both the spinor and vectorial indices. Importantly, the space-time amplitude of the two-particle evolved state (\ref{Psi-coor-1}) is a functional of the customary plane-wave matrix element $S_{fi}^{(1)}$.
The case when \textit{only one} of the final particles is detected is considered in~Refs.~\cite{KBGSS-2022, KBGSS-2023} .

 \begin{figure}\center

\includegraphics[scale=0.6]{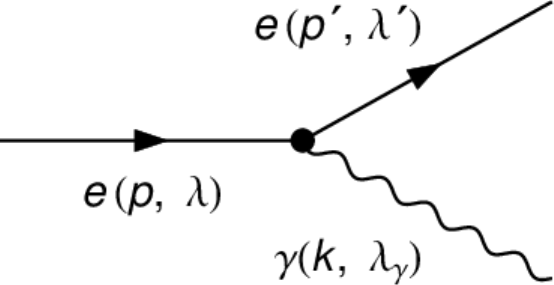}

    \caption{Feynman diagram for the Vavilov--Cherenkov process. The initial electron has a four-momentum $p=(E, \bp)$ and helicity $\lambda$, the final electron is characterised by $p'=(E', \bp')$ and $\lambda'$, and the radiated photon by $k=(\omega, \bk)$ and $\lambda_\gamma$.}
    \label{Fig_diag}
\end{figure}

\subsection{Matrix element of the process}
We proceed with details of the VC radiation by an electron of mass $m_e$ in a transparent nonmagnetic homogeneous medium with the refraction index $ n = n (\omega) $ within the lowest order of QED. This process looks like a decay of an electron (see Fig.~\ref{Fig_diag}) \be
 e(p, \lambda)\to e(p',\lambda')+ \gamma(k, \lambda_\gamma).
 \ee
where for definiteness we have indicated  momenta and helicities of the particles.
In the plane-wave amplitude below, the initial state is described by an electron plane wave $u_{\bp \lambda}\, e^{-ipx}$ with momentum $\bp$, energy $E=\sqrt{\bp^2+m_e^2}$, and helicity $\lambda=\pm 1/2$. The final state is described by a plane wave $u_{\bp' \lambda'}\, e^{-ip'x}$ and a photon plane wave ${\bf e}_{\bk \lambda_\gamma}\, e^{-ik x}$, where the photon frequency is $\omega=|\bk|/n$ and the helicity is $\lambda_\gamma=\pm 1$. 

However, we bear in mind that we intend to use the amplitude calculated here for the process in which initial electron is not exclusively described by a plane wave, and its movement is not necessarily a propagation along a specific axis. Thus, in what follows we keep the spherical angles $\theta,\,\varphi$ and $\theta',\,\varphi'$ of the initial and final electrons and the spherical angles $\theta_\gamma,\,\varphi_\gamma$ of the final photon. An apparent dependance  of the spinors $u\equiv u_{\bp\lambda}$ and $u'\equiv u_{\bp'\lambda'}$ as well as the polarization vector ${\bf e}\equiv {\bf e}_{\bk \lambda_\gamma}$ on the corresponding spherical angles can be given in terms of an expansion over the eigenstates 
of the spin operator $\hat s_z$, $\hat s'_z$ and $\hat s^\gamma_z$. See the particularities in Appendix~\ref{App_tw_properties}. That representation is convenient when we discuss both the plane wave and twisted states.

In the process discussed, the plane-wave matrix element reads
\bea
 & S^{(1)}_{fi} = i (2\pi)^4\,N\, \delta(p'+k-p)\, M_{fi},
 \cr
& M_{fi} = M_{fi}(p,p',k)=\sqrt{4\pi \alpha}\, \;\bar{u}'\,
\gamma_\mu  u\,(e^\mu)^*,
\label{SPW}
 \eea
where the normalization factor is $N=\sqrt{4\pi/(2E 2E' 2\omega n^2)}$ (here we set the factor representing the phase space volume to be equal to unity for brevity).  {We express the standard  amplitude $M_{fi}$ using Eq.~(A5) from Ref.~\cite{ISZ-2016}): 
 \bea
\label{Mfist}
 M_{fi} = \sum_{\sigma=\pm 1/2}\,\sum_{\sigma_\gamma=\pm 1,\,0} e^{i\sigma (\varphi'-\varphi)-i\sigma_\gamma(\varphi'-\varphi_\gamma)}\\ \nn \times
 M^{\lambda\lambda' \lambda_\gamma}_{\sigma \sigma_\gamma} (E,\omega,\theta, \theta',\theta_\gamma),
 \eea
where the dependence on the azimuthal angles $\varphi, \varphi',\varphi_{\gamma}$ is factorized and the coefficients
  \bea \label{Mlong}
M^{\lambda\lambda' \lambda_\gamma}_{\sigma \sigma_\gamma}&=&
-\sqrt{4\pi \alpha}\,  2\sigma \,2\lambda \,
E_{\lambda \lambda'}\,d^{1/2}_{\sigma \lambda}(\theta)
 \\
 &\times&  d^{1/2}_{\sigma-\sigma_\gamma, \lambda'}(\theta')\,
 d^{1}_{\sigma_\gamma,\,\lambda_\gamma}(\theta_\gamma)\,
   \left(\delta_{0,\, \sigma_\gamma}
 -\sqrt{2}\,\delta_{2\sigma,\sigma_\gamma}\right)
 \nn
 \eea
do not depend on these angles. Here
 \bea
  \label{Ell}
 E_{\lambda \lambda'}&=&\sqrt{(E-m_e)(E'+m_e)}
 \\
 &+&2\lambda 2\lambda' \sqrt{(E'-m_e)(E+m_e)}
 \nn
 \eea
 }
and $d_{MM'}^{\;\;\;\,J}\left(\theta\right)$ are the small Wigner matrices~\cite{Varshalovich}:
 \bea
 \label{Wigner}
d^{\,1/2}_{\sigma \lambda}(\theta)&=&\delta_{\sigma \lambda}
 \cos(\theta/2)-2\sigma \delta_{\sigma,- \lambda} \sin(\theta/2),
 \\
 d^{\;1}_{0,\lambda_\gamma}(\theta_\gamma)&=&
 \fr{\lambda_\gamma}{\sqrt{2}}\,\sin\theta_\gamma,\;\;
 d^{\;1}_{2\sigma,\,\lambda_\gamma}(\theta_\gamma)=
 \frac{1}{2}\,\left(1+ 2\sigma \lambda_\gamma \cos\theta_\gamma \right).
 \nn
 \eea

{It is useful to point out the following feature of  the discussed process: 
the angle $\theta_{kp}$ between vectors $\bk$ and ${\bp}$ is completely determined by the energies of the photon and the initial electron. Indeed, if we rewrite the equality $E' = E-\omega$ in the form
 \be
(E')^2=\bp^2+\bk^2- 2 |\bp| |\bk| \cos\theta_{kp} + m_e^2=(E-\omega)^2,
 \ee
we get  (cf. \eqref{1})
}
 \be 
 \cos\theta_{kp} =  \fr{1}{vn}+\fr{\omega}{2E}\,\fr{n^2-1}{vn}.
 \label{thetapk}
 \ee 
 
\section{Plane-wave initial electron}
\label{Sec_PW}
\subsection{Evolved state}
\begin{figure}\centering 
\includegraphics[scale=0.45] {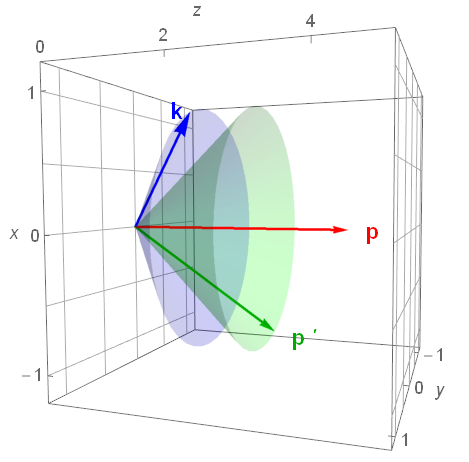} \qquad
\includegraphics[scale=0.35] {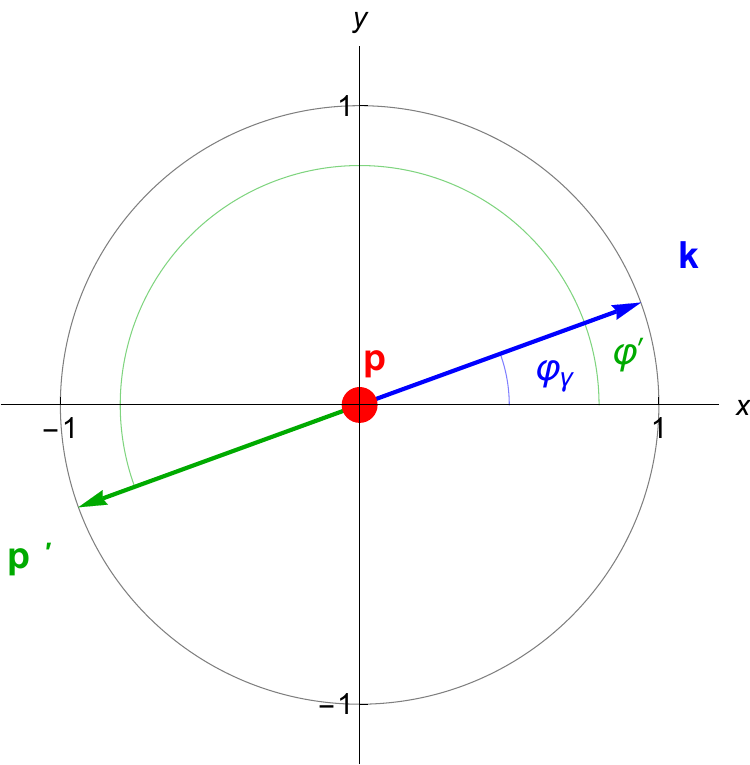}
    \caption{The geometry of the VC process with the initial electron moving along the $z$-direction.}
    \label{Fig_cone_alongz}
\end{figure}
Let us suppose that the incoming electron has a definite momentum, directed along the $z$-axis, so that $\theta=\varphi=0$, $p=(E, 0, 0 , |\bp|)$ as  depicted in the top panel of Fig.~\ref{Fig_cone_alongz}. { Therefore, from~\eqref{thetapk} we immediately obtain a characteristic equality
  \be
\theta_\gamma = \theta_{kp} = \theta_0,
\label{gamma_gamma0}
  \ee
which means that {the VC radiation with a definite energy $\omega$ is concentrated on the cone with the opening angle} $\theta_0$.
}

In that case we deal with an electron that can be described standardly as a plane wave with $\hat j_z =\hat s_z$ (see Appendix \ref{App_tw_properties}). For the matrix element, this results in the reduction $d^{\,1/2}_{\sigma \lambda}(\theta=0)=\delta_{\sigma \lambda}$ and, hence, we can lift both sums in Eq.~(\ref{Mfist}) 
 \be
\label{Mfist_s}
 M_{fi} =  M^{\lambda\lambda' \lambda_\gamma}_{\lambda, 0}\,
 e^{i\lambda \varphi'}+ 
  M^{\lambda\lambda' \lambda_\gamma}_{\lambda, 2\lambda}\,
 e^{-i\lambda \varphi'+i2\lambda \varphi_\gamma},
 \ee
where the new coefficients are {
  \bea
   \label{Mfist_s1}
&&M^{\lambda\lambda' \lambda_\gamma}_{\lambda, 0}=
-\sqrt{4\pi\alpha} E_{\lambda \lambda' }\,
 d^{1/2}_{\lambda, \lambda'}(\theta')\,
 d^{1}_{0,\,\lambda_\gamma}(\theta_\gamma),
 \nn
 \\ 
 \label{Mfist_s2}
&&M^{\lambda\lambda' \lambda_\gamma}_{\lambda, 2\lambda}=
+\sqrt{8\pi\alpha} E_{\lambda \lambda' }\,
 d^{1/2}_{-\lambda, \lambda'}(\theta')\,
 d^{1}_{2\lambda,\,\lambda_\gamma}(\theta_\gamma).
 \nn
 \eea
}
The evolved state in the space-time representation, given by Eq.~\eqref{Psi-coor-1},
defines the probability amplitude to detect the photon in a region of space-time centered in the point $x_{\gamma} = (t_{\gamma},{\bm r}_{\gamma})$ while the electron is jointly detected in a region centered in the point $x_e = (t_e,{\bm r}_e)$ (both the points may or may not coincide). 
First, we integrate over the three-momentum of the final electron and then integrate over the polar angle of the photon $\theta_\gamma$ rendering it to the value defined by Eq.~(\ref{thetapk}). 
 This leads to the replacement
 \bea \label{Gamma_plw}
 d\Gamma =&  (2\pi)^4 \delta(p'+k-p)\,\fr{d^3 p'}{(2\pi)^3}\fr{d^3 k}{(2\pi)^3}\\ &\rightarrow\ \fr{(E-\omega)\omega n(\omega)}{v E}\, \fr{d(\omega n(\omega))}{2\pi}\,
 \fr{d\varphi_\gamma}{2\pi}. \nn
 \eea
 When the dispersion is weak, $|\frac{\omega}{n}\frac{dn}{d\omega}| \ll 1$, we can have $d(\omega n(\omega)) \approx  n(\omega)d\omega$ and this approximation is used in some formulas below. 
 The evolved state looks as follows:
  \begin{align} \label{EvX1}
\displaystyle \Psi^{\rm (pw)}_{\bp \lambda}(x_e, x_\gamma) &\equiv \la 0| \hat{\psi}(x_e) \hat{{\bf A}}(x_{\gamma})|e',\gamma\ra^{(\text{ev})}
\\  & \displaystyle =  \frac {i} {v(2E)^{3/2}}\,
\int \frac{d(\omega n)}{n}\, e^{-i\omega t_{\gamma} - i (E-\omega)t_e} \cr 
 & \displaystyle \times \sum_{\lambda' \lambda_\gamma }\int \fr{d\varphi_\gamma}{2\pi}\, u_{\bp'\lambda'}\, e^{i(\bp - \bk)\cdot {\bm r}_e}\, {\bm e}_{\bk \lambda_{\gamma}}\, e^{i \bk\cdot {\bm r}_{\gamma}} M_{fi}.
\nn
\end{align}
Here we have used the following four-vectors 
 \bea
 p'=(E',\bp')=(E-\omega, \bp-\bk),\\  k=(\omega, k_\perp \cos\varphi_\gamma,k_\perp \sin\varphi_\gamma, \omega n \cos\theta_{kp}),\nn
  \eea
where $k_\perp =\omega n \sin\theta_{kp}$ and $\theta_{kp} = \theta_{kp}(E,\omega)$ is defined by Eq.~\eqref{thetapk}. Moreover, in the used reference frame we have $\bp'_\perp=-\bk_\perp$ and, therefore, only two points give contribution to the integral over $\varphi_\gamma$:
 \bea \label{phi}
 &\varphi' = \varphi_\gamma+ \pi \;\; \mbox{when}\;\; \varphi_\gamma \subset (0; \pi)\\ \nn
 \mbox{or}\;\;
&\varphi'=\varphi_\gamma-\pi \;\; \mbox{when}\;\; \varphi_\gamma \subset (\pi; 2\pi), 
 \eea
the first option is presented in the bottom panel of Fig.~\ref{Fig_cone_alongz}, and the second is easy to imagine with  $\bp'$ and $\bk$ swapping places ($\varphi_\gamma $ becoming greater than $180^\circ$ with $\varphi'$ remaining in $(0; 2\pi)$ domain).
Therefore
 \bea
\label{Mat}
 M_{fi}|_{\bp'_\perp=-\bk_\perp} =  e^{i\lambda (\varphi_{\gamma} \pm \pi)}\left (M^{\lambda\lambda' \lambda_\gamma}_{\lambda, 0} - 
  M^{\lambda\lambda' \lambda_\gamma}_{\lambda, 2\lambda}\right ).
 \eea

Now, let us expand the momentum states of the final particles in Eq.~(\ref{EvX1}) over the complete sets of the {\it twisted states} \cite{Knyazev-Serbo-2018}:
\bea
& \displaystyle u_{\bp'\lambda'}\, e^{-ip'x_e} = \sum\limits_{m' = -\infty}^{+\infty} i^{m'}\, e^{-im'\varphi'} \psi_{p'_{\perp}p'_z m' \lambda'} (x_e), \label{series_spinor} \\
& \displaystyle
{\bm e}_{\bk \lambda_{\gamma}}\, e^{-i k x_{\gamma}} = \sum\limits_{m_{\gamma} = -\infty}^{+\infty} i^{m_{\gamma}} e^{-im_{\gamma}\varphi_{\gamma}} {\bm A}_{k_{\perp}k_z m_{\gamma} \lambda_{\gamma}} (x_{\gamma}) \label{series_vector}
\eea
where $m'$ is a half-integer, $m_{\gamma}$ is an integer, and the functions  ${\bf A}_{k_\perp k_z m_\gamma \lambda_\gamma}(x_\gamma)$ and $ \psi_{p'_\perp  p'_z m' \lambda'}(x_e)$ are defined in Appendix~\ref{App_tw_properties}. After this, the azimuthal integral in Eq.~(\ref{EvX1}) yields $\delta_{\lambda,m'+m_{\gamma}}$ revealing the conservation of the angular momentum projection. The final result can be presented in the following way:
\bea 
\nn
 &\displaystyle \Psi^{\rm (pw)}_{\bp \lambda}(x_e, x_\gamma) =  \frac {i^{\lambda+1}} {v(2E)^{3/2}}\,
\sum_{\lambda' \lambda_\gamma } \sum\limits_{m', m_\gamma = -\infty}^{+\infty} \int \frac{d(\omega n)}{n}\, 
\\ \nn
 & (-1)^{ m_{\gamma}} \left(M^{\lambda\lambda' \lambda_\gamma}_{\lambda, 0} - 
  M^{\lambda\lambda' \lambda_\gamma}_{\lambda, 2\lambda}\right)\,\delta_{\lambda,m'+m_\gamma}\\
  &\times {\bm A}_{k_{\perp}k_z m_{\gamma} \lambda_{\gamma}} (x_{\gamma})\, \psi_{p'_{\perp}p'_z m'\lambda'} (x_e)
  \label{EV2-1}
\eea
where $E'=E-\omega,\, p'_{\perp} = k_{\perp} = \omega n \sin \theta_{kp},\, p'_z = p-k_z$.The integration is performed over the region defined by the energy conservation law $0<\omega < E-m_e$ and the condition $0< \cos\theta_{kp} < 1$. 

Let us take a closer look at Eq.~\eqref{EV2-1}: it contains the function ${\bf A}_{k_\perp k_z m_\gamma  \lambda}(x_\gamma)$, which describes a twisted photon with the corresponding total angular momentum (TAM) $\hat j^{\gamma}_z$ equal to $m_\gamma$ and certain values of $k_\perp$, $k_z$ and $\lambda_\gamma$; there is also the function $\psi_{p'_\perp p'_z,\lambda-m_\gamma,\lambda'}(x_e)$, which describes a twisted electron with $\hat j'_z$  equal $\lambda-m_\gamma$, and certain values of $p'_\perp$, $p'_z$ and $\lambda'$.  
Therefore, the evolved wave function of the final system \eqref{EV2-1} as a whole describes an {\it entangled state} of a twisted electron and twisted photon.
{If  we  define the TAM operator of the evolved state as a  sum of TAM operators, $\hat J_z=\hat j'_z+\hat j^\gamma_z$, the function $\Psi^{\rm (pw)}_{\bp \lambda}(x_e, x_\gamma)$ is the eigenstate of $\hat J_z$ with the eigenvalue $\lambda$. This result is in accordance with the TAM conservation law   $\hat j'_z+\hat j^\gamma_z=\hat j_z$, since the initial state in the considered case  is an eigenfunction of the TAM operator $\hat j_z=\hat s_z$ having the eigenvalue  $\lambda$. }

Alternatively, we can obtain the evolved wave function in the momentum representation in the form
  \bea 
  \label{phi-st}
\Phi^{(\rm pw)}_{\bp \lambda}(\bp', \bk)&=&
\sum\limits_{\lambda' \lambda_{\gamma}}  u_{{\bf p}'\lambda'}\, {\bf e}_{{\bf  k}\lambda_{\gamma}}\, S_{fi}^{(1)}
 \\
&=& i (2\pi)^4 \delta(p'+k-p) N e^{i\lambda (\varphi_{\gamma} \pm \pi)} \nn \\
&&\sum\limits_{\lambda'\lambda_{\gamma}}  u_{{\bf p}'\lambda'}\, {\bf e}_{{\bf  k}\lambda_{\gamma}}\, 
  \left (M^{\lambda\lambda' \lambda_\gamma}_{\lambda, 0} - 
  M^{\lambda\lambda' \lambda_\gamma}_{\lambda, 2\lambda}\right ).
  \nn 
  \eea 
The explicit expression for this helicity sum is given separately in Appendix~\ref{App_helicity_sum}.
Certainly, the evolved wave-function~\eqref{phi-st} also is the eigenfunction of  $\hat J_z$  with the same eigenvalue $\lambda$.

\subsection{ Soft photon approximation}
In the commonly used detectors of the Cherenkov radiation the photon energy $\omega$ is about several eV and the conditions $\omega \ll m_e $ and $ \omega \ll E-m_e $ are satisfied. In this {\it soft photon approximation} several simplifications to the calculation of the evolved wave function can be made. As $E' \approx E$, we get $\delta_{\lambda' \lambda}$ from Eq.~\eqref{Mlong}. Eventually, with $\bp' \approx \bp$ and $\theta'$ set to zero, only one term remains from the  Wigner function multiplication and
 the expression for the scattering amplitude is 
 \be
 M_{fi}=-\lambda_\gamma \sqrt{8\pi \alpha}\, e^{i\lambda (\varphi_{\gamma} \pm \pi)} \,vE \sin\theta_0\delta_{\lambda' \lambda},
 \ee
where
 \be
 \sin\theta_0=\fr{k_\perp}{|\bk|}=\sqrt{1-\fr{1}{v^2n^2}}.
 \ee
The summation over $\lambda'$ and $\lambda_\gamma$ in \eqref{phi-st} is performed as detailed in Appendix~\ref{App_helicity_sum}, and the resulting evolved wave function is greatly simplified when written in terms of the longitudinal photon polarization ${\bf e}_{\parallel}$ (see Eq.~\eqref{paral} in Appendix~\ref{App_tw_properties})
\bea
 \Phi^{\rm (pw)}_{\bp \lambda}({\bf p}^\prime,\,{\bf k}) &=&
 i4\sqrt{\pi\alpha} (2\pi)^4 \delta(p'+k-p)\,\\
 \nn && \times N v E\, \sin\theta_0 \,
 u_{\bp' \lambda}\,{\bf e}_{\parallel}\,e^{i\lambda (\varphi_{\gamma} \pm \pi)}
  \eea
in the momentum representation and
 \bea
 \label{Psi-app}
 \Psi^{\rm (pw)}_{\bp \lambda}(x_e, x_\gamma)&=& i\,  \sqrt{\fr{2\pi\alpha}{E}} \,
 u_{\bp \lambda}\, e^{-i p x_e}\\
 \nn
 &&\times
  \int  \sin\theta_0\,
 {\bf A}_{k_\perp, k_z,0, \parallel}(x_\gamma)\, d\omega
   \eea
in the coordinate representation, where the photon wave function with the longitudinal polarization is given in Eq.~\eqref{bessel-wave-l} and the region of integration over $\omega$ is determined by the condition $v n(\omega) > 1$.  The obtained result corresponds to the linear polarization in the plane of the vectors $ \bp $ and $ \bk $. This fact is well known from the first experiments of Cherenkov.

This result also shows that the value $ m_\gamma = 0 $ is consistent with the conservation law for the $ z $-projection of the system TAM,  $ \lambda = \lambda'+ m_\gamma $, since in the approximations used, the helicity state and momentum of the final electron almost coincide with those of the initial electron, i.\,e.
$ \lambda'= \lambda,\;\;p'=p $.

\subsection{Approximation of ultra-relativistic electrons}

The amplitude of the process could also be considered in the {\it approximation of ultra-relativistic electrons}, implying that $E\gg m_e$ and $E'\gg m_e$:
 \bea
 \label{D}
M^{\lambda\lambda' \lambda_\gamma}_{\sigma \sigma_\gamma}&=&-\sqrt{4\pi \alpha}\,
 8\sigma\lambda\delta_{\lambda \lambda'} \sqrt{EE'}
 \, d^{\,1/2}_{\sigma \lambda}(\theta)\,
 \\
 \nn
 &&d^{\,1/2}_{\sigma-\sigma_\gamma, \lambda}(\theta')\,d^{\;1}_{\sigma_\gamma,\,\lambda_\gamma}(\theta_\gamma)\,
   \left(\delta_{0,\, \sigma_\gamma}
 -\sqrt{2}\,\delta_{2\sigma,\sigma_\gamma}\right).
 \eea
In this limit, helicities of electrons are conserved and summation over $\lambda'$ becomes trivial due to the appearance of $\delta_{\lambda\lambda'}$.  Indeed, using relations \eqref{Mat}, we obtain 
{ 
\be
M_{fi}= -\sqrt{8\pi \alpha  E E'}\, \delta_{\lambda \lambda'} \, e^{i\lambda (\varphi_{\gamma} \pm \pi)}
\left( \lambda_\gamma g_1 + g_2\right)\,,
 \ee
where
 \be \label{pw_g_def}
 g_1 =\sin(\theta_\gamma+ \mbox{$\frac 12$}\theta'),\;\;
 g_2=2\lambda \sin(\mbox{$\frac 12$}\theta').
 \ee
Substituting this expression into \eqref{phi-st} and performing
summation over $\lambda_\gamma$  with the help of \eqref{e-to-e}, we obtain the result in terms of the twisted photon states with the linear polarization  for the evolved wave function in the momentum representation (see Appendix~\ref{App_helicity_sum})
 \bea
  \Phi^{\rm (pw)}_{\bp \lambda}({\bf p}^\prime,\,{\bf k})& =&
-4i\,  (2\pi)^4 \delta(p'+k-p) N\,
\\\nn
&\times& \,
\sqrt{\pi\alpha E E'} u_{\bp' \lambda}  e^{i\lambda (\varphi_{\gamma} \pm \pi)}
\left( g_1\, {\bf e}_{\parallel}+i g_2\,{\bf e}_{\perp}\right)
 \nn
 \eea
and in the coordinate representation
 \bea
 \label{Psi-pl-tw}
&\displaystyle\Psi^{\rm (pw)}_{\bp \lambda}(x_e, x_\gamma)= \frac{i^{\lambda+1}}v \sqrt{\fr{2\pi \alpha}{E}}
   \sum_{m_\gamma} \int  d\omega \sqrt{1-\omega/E}
   \\\nn
   & \times (-1)^{m_\gamma}    \psi_{p'_\perp,  p'_z,\lambda-m_\gamma, \lambda}(x_e)
   \\
  &\times  \bigg[g_1 {\bf A}_{k_\perp k_z m_\gamma,\parallel}(x_\gamma) +i\, g_2  {\bf A}_{k_\perp k_z m_\gamma,  \perp}(x_\gamma)\bigg]\,.
    \nn
 \eea

We observe that the wave function of the final state with the longitudinal linear polarization (i.e., lying in the same plane as the vectors $ \bp $ and $ \bk $) is proportional to $g_1$ while the function with the linear polarization orthogonal to the above-mentioned plane is proportional to  $i g_2$. Therefore, we can construct a quantity with the meaning of the degree of linear polarization of the final photon
 \be \label{degreePldef}
 P^{\rm (pw)}_l = \fr{|g_1|^2 -|g_2|^2}{|g_1|^2 +|g_2|^2}
 \ee
for any value of $m_\gamma$. Under the standard conditions for VC radiation, i.e. when $\theta' \ll \theta_\gamma$ and
$\omega \ll E$, we have
 \be
 P^{\rm (pw)}_l =1- d,\;\; d = \fr{\omega^2 (n^2-1)}{2(E\sin\theta_0)^2} \ll 1.
 \label{degreePlpw}
 \ee
This result coincides with the one given for the degree of linear polarization of emitted VC radiation in Ref.~\cite{ISZ-2016} (see Eq. (15) there). Besides, it broadens the applicability of the latter
, in particular, it demonstrates that the expression is valid for any values of $m_\gamma$.
It can also be seen from Eq.~\eqref{degreePlpw} that the longitudinal linear polarization is dominating in the region where $\theta'\ll \theta_\gamma$. 
}
 
\section{Twisted initial electron}
\label{Sec_TW}
If the initial electron is in {\it the twisted state}, then its plane-wave function $u\, e^{-ip x}$ is replaced by a superposition of the plane waves 
  \be
 \label{electron-bessel-wave-evident}
 u_{\bp \lambda}\, e^{-ip x} \to
 \int \fr{d\varphi}{2\pi}\, i^{-m} \,e^{im\varphi}\,u_{\bp \lambda}\, e^{-ip x}.
  \ee
This is the stationary state ({\it the Bessel wave}) corresponding to a twisted electron with the longitudinal momentum $ p_z $, the absolute value of transverse momentum  $ p_\perp $, the energy $ E = \sqrt {p^2_\perp + p_z^2 + m_e^2} $, the projection of an electron TAM onto the $ z $-axis
equal to a half-integer number $m$ and the helicity $ \lambda = \pm 1/2 $.
In this case, the evolved wave function of the final state  has the form
\bea
\label{phi-tw}
 \Phi^{\rm (tw)}_{p_\perp p_z  m \lambda}(\bp',\bk)&=&
 \int \fr{d\varphi}{2\pi}\, i^{-m} \,e^{im\varphi}  \,\Phi^{\rm (pw)}_{\bp\lambda}(\bp',\bk)
   \eea
in the momentum representation and the similar expression in the coordinate representation. Note, that the function $\Phi^{\rm (pw)}_{\bp\lambda}(\bp',\bk)$ in the wave function of Eq.~\eqref{phi-tw}  depends now on $M_{fi}$ in the form given by Eq.~\eqref{Mfist} and bears all three components of the vector 
$\bp=(p_\perp\cos\varphi,\,p_\perp\sin\varphi,\, p_z)$ (as for example in Fig.~\ref{Fig_cone}), while we took the frame with $z$-directed $\bp=(0,\,0,\,|\bp|)$ when deriving $\Phi^{\rm (pw)}_{\bp\lambda}(\bp',\bk)$ in Eq.~\eqref{phi-st}. 

\subsubsection{Integration over the azimuth of initial particle}
\begin{figure}\center
\includegraphics[scale=0.45] {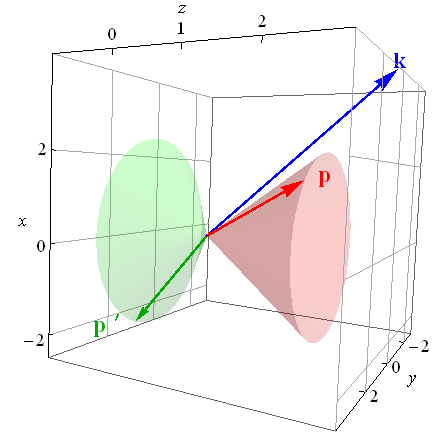}
\ \quad \ 
\includegraphics[scale=0.45] {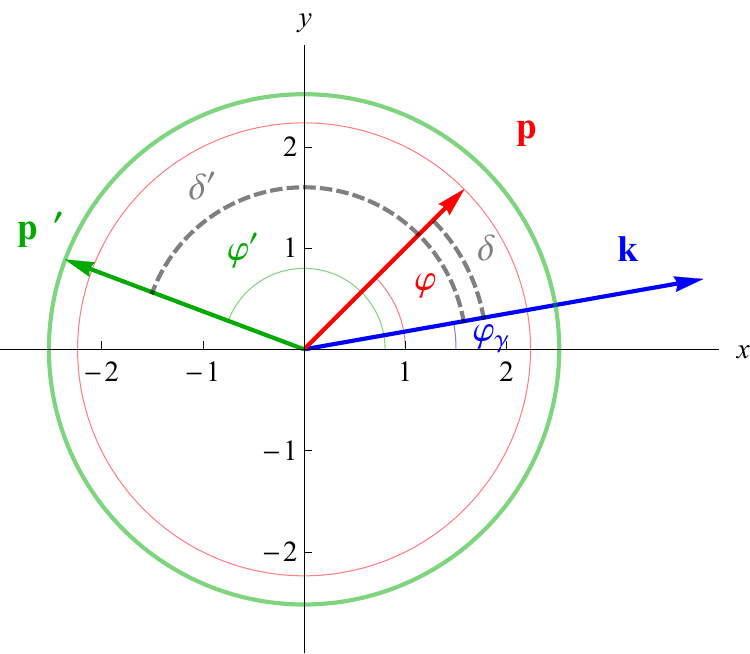}
    \caption{The geometry in the process with initial twisted electron having non-zero $p_\bot$. The lengths and directions of the vectors are illustrative but not exactly natural for the process.}
    \label{Fig_cone}
\end{figure}
To get the final state in either representation the integration over $\varphi$ can be performed first. We must take into account that $\cos\theta_{kp}$ generally depends on $\varphi$ and also that the plane-wave integrand has some dependence on $\varphi$ and $\varphi'$ -- let us call it $f(\varphi, \varphi')$. The following integral arises
 \be \label{az_integral}
I=\int_0^{2\pi} \fr{d\varphi}{2\pi}  \delta\left( \cos\theta_{kp}-\cos\theta_0 \right)
f(\varphi, \varphi'),
 \ee
 where $\theta_0 = \arccos\left [\fr{1}{vn}+\fr{\omega}{2E}\,\fr{n^2-1}{vn}\right]$  is the value determined by the conservation law (recall Eq.~\eqref{thetapk}).
The integral is of the same type as the one studied in Ref.~\cite{ISZ-2016}. To calculate it we follow the same route with appropriate extension  treating the appearance of $\varphi'$. 

It is straightforward to get the following equations for the angles between vectors $(\bk, \bp)$ or $(\bk, \bp')$ if the definitions $\cos\theta_{kp}=\fr{\bk\cdot\bp}{|\bk|\,|\bp|}$ 
and $\cos\theta_{kp'}=\fr{\bk\cdot\bp'}{|\bk|\,|\bp'|}$ are written in spherical coordinates:
\bea \nn
&\cos\theta_{kp}=\cos \theta_\gamma \cos\theta+\sin \theta_\gamma \sin\theta \cos \delta, \\ 
&\cos\theta_{kp'}=\cos \theta_\gamma \cos\theta'+\sin \theta_\gamma \sin\theta' \cos \delta',\\ \nn
&\delta = \pm (\varphi-\varphi_\gamma), \ \delta' = \pm (\varphi'-\varphi_\gamma).
\eea
Further, since $\cos\theta_{kp}$ is fixated by the delta function, $\varphi$ could be expressed other way round with $\delta$, while for $\varphi'$ there could be only one option as is evident from the geometrical construction akin to that in the bottom panel of Fig.~\ref{Fig_cone}. In the end, only two points give nonzero contribution to the integral \eqref{az_integral}: $\varphi=\varphi_\gamma+\delta,\;
\varphi'=\varphi_\gamma+\delta'$ and $\varphi=\varphi_\gamma-\delta,\;
\varphi'=\varphi_\gamma-\delta'$  (see Fig.~\ref{Fig_cone}), where
 \bea
 \label{delta}
  \delta&=& \arccos\left(\fr{\cos\theta_0 - \cos\theta_\gamma \cos\theta}
 {\sin\theta_\gamma \sin\theta} \right),\\
 \label{delta'}
  \delta'&=& \arccos\left(\fr{\cos\theta_{kp'} - \cos\theta_\gamma \cos\theta'}
 {\sin\theta_\gamma \sin\theta'} \right).
 \eea
The final expression is
\be 
I= \fr 12 \left[f(\varphi_\gamma+\delta, \varphi_\gamma+\delta')+
f(\varphi_\gamma-\delta,\, \varphi_\gamma-\delta') \right]
F(\theta, \theta_\gamma, \theta_0)
 \ee 
 where the function  $F(\theta, \theta_\gamma, \theta_0)$ reads
  \bea 
   \label{F}
 &F(\theta, \theta_\gamma, \theta_0)= \fr{1}{\pi\,\sin\theta_\gamma \sin\theta |\sin\delta|}
  \\
 &= \fr{1}{\pi}\;
 \left\{ \left[\cos\theta_\gamma -\cos(\theta+\theta_0)\right]
 \left[\cos(\theta-\theta_0)-\cos\theta_\gamma\right] \right\}^{-1/2}
 \nn
  \eea
\subsubsection{Momentum representation}
To calculate the evolved state of Eq.~\eqref{phi-tw}, we collect the azimuthal dependencies into
 \be
 f(\varphi, \varphi')=i^{-m}e^{im\varphi-i\sigma'\varphi'+i\sigma (\varphi'-\varphi)-i\sigma_\gamma(\varphi'-\varphi_\gamma)}.
 \ee 
 and the resulting expression for \eqref{az_integral} is
   \bea
I&=& i^{-m} e^{i(m-\sigma')\varphi_\gamma}\\ \nn
&\times& F(\theta, \theta_\gamma, \theta_0) \, \cos\left[(m-\sigma)\delta + (\sigma-\sigma'-\sigma_\gamma)\delta'\right],
 \eea

Then, the evolved wave function in the momentum representation is
  \bea 
&&\Phi^{(\rm tw)}_{p_\perp p_z m \lambda}(\bp', \bk)=
 i^{1-m} (2\pi)^4 \delta(\bp'+\bk-\bp) 
   \label{phi-tw-final} \\ \nn
 &&\times \fr{(E-\omega) N}{v E \omega n} F(\theta, \theta_\gamma, \theta_0) \\
 \nn
 &&\,\sum\limits_{\lambda' \lambda_{\gamma}} \sum_{\sigma\sigma'\sigma_\gamma} 
 e^{i(m-\sigma')\varphi_\gamma}
\,d^{(1/2)}_{\sigma' \lambda'}(\theta')M^{\lambda\lambda' \lambda_\gamma}_{\sigma \sigma_\gamma}(E,\omega,\theta, \theta',\theta_\gamma)
\\
&&\times  
 \cos\left[(m-\sigma)\delta +
(\sigma-\sigma'-\sigma_\gamma)\delta'\right] {\bf e}_{\bk \lambda_\gamma} U^{(\sigma')}(E',\lambda') ,\nn
\eea 
where we give a full expression in terms of bispinors $U^{(\sigma)}$ defined in Eq.~\eqref{eq_U_spinor} and
 $M^{\lambda\lambda' \lambda_\gamma}_{\sigma \sigma_\gamma}$ is taken from  Eq.~\eqref{Mlong}. 
\subsubsection{Two-particle wave function in coordinate representation}
Let us write down explicitly the definition of the evolved state in the coordinate representation with additional integration over $\varphi$ signifying the vortex character of the incoming electron:
  \bea
&& \displaystyle \Psi^{\rm (tw)}_{p_\perp  p_z m\lambda}(x_e, x_\gamma)\equiv\la 0| \hat{\psi}(x_e) \hat{{\bf A}}(x_{\gamma})|e',\gamma\ra^{(\text{ev})} \\ \nn
&& \displaystyle = \int\frac{d^3p'}{(2\pi)^3}\frac{d^3k}{(2\pi)^3}\,\sum\limits_{\lambda' \lambda_{\gamma}}
\fr{u_{{\bf p}'\lambda'}}{\sqrt{2 E'}}\,e^{-ip'x_e}\,
\fr{\sqrt{4\pi}\,{\bf e}_{{\bf k}\lambda_{\gamma}} }{\sqrt{2\omega} n}\, e^{-i kx_\gamma}\;\\
&& \times\int \fr{d\varphi}{2\pi} i^{-m} e^{im\varphi}
 S_{fi}^{(1)} \nn
 \eea
Using again the series expansions \eqref{series_spinor}, \eqref{series_vector} we collect all azimuthal exponents into
 \bea
 f(\varphi, \varphi')&=&e^{im\varphi-im'\varphi'-im_\gamma\varphi_\gamma+i\sigma (\varphi'-\varphi)-i\sigma_\gamma(\varphi'-\varphi_\gamma)}
 \nn\\
 &\times& i^{-m+m'+m_\gamma}.
 \eea
The  integral \eqref{az_integral} results in
 \bea
\displaystyle I &=&i^{-m+m'+m_\gamma}e^{i(m-m'-m_\gamma)\varphi_\gamma}\\ \nn
&\times& \cos[(m-\sigma)\delta+(\sigma-m'-\sigma_\gamma)\delta'] F(\theta, \theta_\gamma, \theta_0).
\eea
The integration in the phase space could be performed with elimination of the $d^3p'$  integral, and we are left with
 \be
d\Gamma \rightarrow \fr{(E-\omega)\omega n^2}{2\pi vE}\,
 \delta\left( \cos\theta_{kp}-\cos\theta_0 \right)
 \frac{d(\omega n)}{n} \sin\theta_\gamma d\theta_\gamma
 \fr{d\varphi_\gamma}{2\pi}
 \ee

The dependence of the wave function  on $\varphi_\gamma$ is trivial and implies appearance of the Kronecker delta $\delta_{m-m'-m_\gamma, 0}$.
{Finally, the evolved wave-function can be expressed as

\bea
  \label{TW_EV2-1}
& \displaystyle \Psi^{\rm (tw)}_{p_\perp  p_z m\lambda}(x_e, x_\gamma)  =  \fr {i} {v(2E)^{3/2}}\,
\sum\limits_{\lambda' \lambda_\gamma m' m_\gamma}\delta_{m, m'+m_\gamma} \\ \nn
& \displaystyle \times
\int \fr{d(\omega n)}{n} d\theta_\gamma \sin\theta_\gamma\,\ F(\theta, \theta_\gamma, \theta_0)\, C_{\lambda' \lambda_\gamma m_\gamma}
 \cr 
 &\times\displaystyle {\bm A}_{k_{\perp}k_z m_{\gamma} \lambda_{\gamma}} (x_{\gamma})\, \psi_{p'_{\perp}p'_z m' \lambda'} (x_e),
\eea
where
\bea \label{coefficient-C}
C_{ \lambda' \lambda_\gamma m_\gamma}&= &\sum_{\sigma \sigma_\gamma} M^{\lambda\lambda' \lambda_\gamma}_{\sigma \sigma_\gamma}(E,\omega,\theta, \theta',\theta_\gamma) \\
&\times&\,\cos[(m-\sigma)(\delta-\delta')+(m_\gamma-\sigma_\gamma) \delta'] .
\nn
\eea 
}
The further integration is performed over the region determined by the energy conservation law $0<\omega < E-m_e$ and the condition~\eqref{interval} discussed below. 
  
It is seen from Eq.~\eqref{TW_EV2-1} as well as from~\eqref{phi-tw-final} that the evolved wave function of the final system is an entangled state of two particles of different type - an electron and a photon, either of which has OAM.
Note, also, that the evolved functions for the twisted initial electron has the same structure as the wave function~\eqref{phi-st} for the plane-wave initial electron except two important points:

{\it (i)}  $\Phi^{(\rm pw)}_{\bp \lambda}(\bp', \bk)$ and $\Psi^{\rm (pw)}_{\bp \lambda}(x_e,x_\gamma)$ being the eigenfunctions of $\hat J_z$ with the eigenvalue $\lambda$ are replaced with $\Phi^{(\rm tw)}_{p_\perp  p_z m \lambda}(\bp', \bk)$ and  $\Psi^{(\rm tw)}_{p_\perp  p_z m\lambda}(x_e,x_\gamma)$, respectively, being the eigenfunctions of $\hat J_z$  with the eigenvalue $m$; 

{\it (ii)} an additional function $F(\theta, \theta_\gamma, \theta_0)$ appears.  This new important function  is the same as in Ref.~\cite{ISZ-2016} (see there the graphs for it) and it is non-zero only when the three theta angles satisfy the ``triangle inequality''
 \be
 \left|\theta - \theta_0 \right| < \theta_\gamma <
 \theta + \theta_0\,.
 \label{interval}
 \ee
Thus, the situation here differs in comparison with the standard VC radiation of the plane wave electron. Namely, the final photon is emitted not along the surface of the cone defined by the angle $ \theta_0 $, but in the region between the two cones given by Eq.~\eqref{interval}. It was proven in Ref.~\cite{ISZ-2016} that, though $F(\theta,\theta_\gamma,\theta_0)$ diverges at the borders of the interval \eqref{interval},  this singularity can be integrated
\be
 \int_{\left|\theta - \theta_0 \right|}^{\theta + \theta_0}
 F(\theta, \theta_\gamma, \theta_0)\,\sin\theta_\gamma\, d\theta_\gamma = 1.
 \label{intF}
  \ee
  
\subsection{Soft photons approximation}
\begin{center}
\begin{figure*}\center
\includegraphics[scale=0.55] {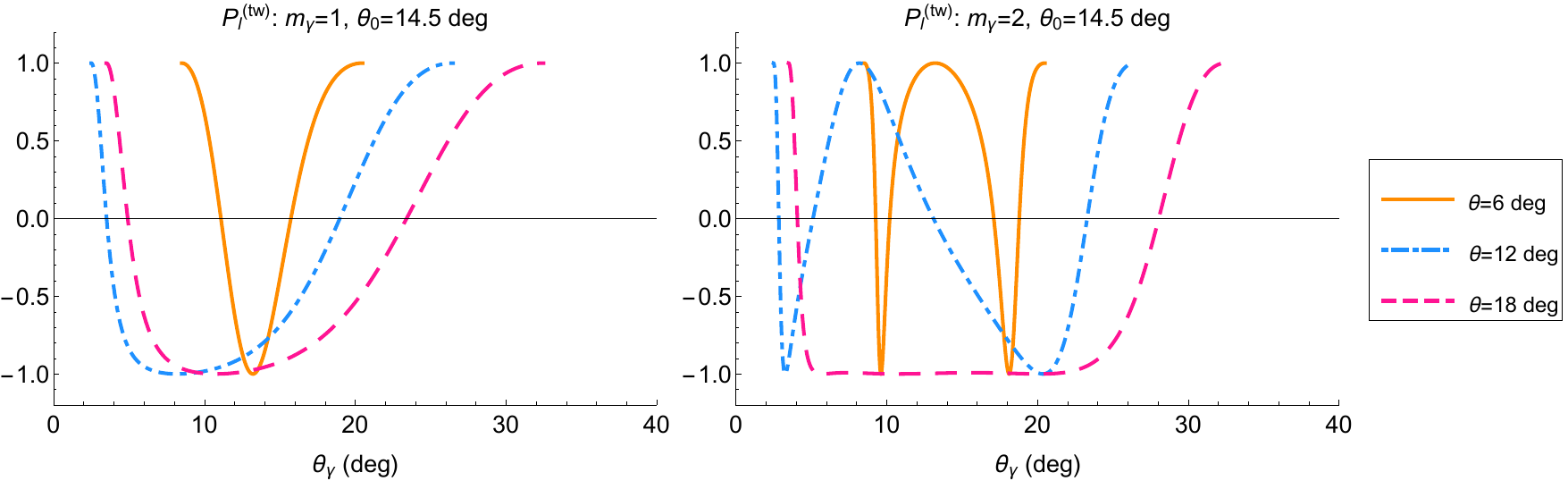}
    \caption{The degree of linear polarization $P^{\rm (tw)}_l$ from Eq.(\ref{degreePltw}) as a function of $\theta_\gamma$ for $\theta_0=14.5 ^{\circ}$ and $\theta=6^{\circ},\ 12^{\circ},\ 18^{\circ}$. The plots occupy the region permitted by Eq.~\eqref{interval}.}
    \label{Fig_polarization_th14}
\end{figure*}
\end{center}

Once again, in the soft photon approximation, the conditions $ \omega \ll m_e $ and $ \omega \ll E-m_e $ are satisfied and the momenta and helicities of the initial and final electrons coincide. In this case we can simplify the corresponding wave-functions with the following relations
 \be
\lambda'=\lambda,\;\; \theta'= \theta,\;\; \delta'=\delta
 \ee
and perform the summation over $\sigma$ and $\sigma_\gamma$.  Let us consider in detail this limit for the evolved wave-function in the coordinate representation. { First, we apply this approximation in the expression for the coefficient in Eq.~\eqref{coefficient-C}:
 \bea \label{coefficient-C-soft}
C_{ \lambda' \lambda_\gamma m_\gamma}&= & -\sqrt{8\pi \alpha}\,vE\, \delta_{\lambda \lambda'} \,
(A+B), \nn
\\
A&=& 2\lambda \sqrt{2} \cos(m_\gamma \delta)\sum_\sigma 2\sigma \left[ d^{1/2}_{\sigma \lambda}(\theta)\right]^2 
d^{\;\;1}_{0, \lambda_\gamma}(\theta_\gamma),
\nn\\ \nn
B&=& -4\lambda \sum_\sigma 2\sigma  d^{1/2}_{\sigma \lambda}(\theta) d^{1/2}_{-\sigma, \lambda}(\theta) 
d^{\;\;1}_{2\sigma, \lambda_\gamma}(\theta_\gamma)\\
\nn
&\times&\cos[(m_\gamma-2\sigma) \delta]
\nn
 \eea 
 }
Using definitions~\eqref{Wigner}
for the Wigner matrices and the relation
\be
\cos[(m_\gamma-2\sigma) \delta]=\cos(m_\gamma\delta) \cos\delta+2\sigma\sin(m_\gamma\delta) \sin\delta,
 \ee
 we find
 \bea \label{G_vortex}
 C_{\lambda' \lambda_\gamma m_\gamma  }&=& -\sqrt{8\pi \alpha}\,vE\, \delta_{\lambda \lambda'}   \,(\lambda_\gamma G_1+ G_2),
 \\
 G_1&=& \left[ \cos\theta\sin\theta_\gamma- \sin\theta\cos\theta_\gamma\cos\delta\right] \, \cos(m_\gamma\delta),\;\;  
  \nn\\
  G_2&=&  -\sin\theta \sin\delta \sin(m_\gamma\delta).
  \nn
  \eea
It is useful to note that Eq.~\eqref{TW_EV2-1}
 can be presented in another form by introducing the linear photon polarizations with the help of Eqs.~\eqref{e-to-e}:
  \bea
\nn
&&\Psi^{\rm (tw)}_{p_\perp p_z  m \lambda}(x_e, x_\gamma)=i\sqrt{\fr{2\pi\alpha}{E}} 
\int \,d\omega \sin\theta_\gamma d\theta_\gamma   F(\theta,\theta_\gamma,\theta_0)\, 
\\
 &&\sum_{m_\gamma} \left[G_1 {\bf A}_{k_\perp k_z m_\gamma ,\parallel}(x_\gamma)
 +iG_2{\bf A}_{k_\perp k_z m_\gamma , \perp}(x_\gamma)\right]
\nn \\
&& \times \psi_{p_\perp p_z, m-m_\gamma, \lambda}(x_e).
  \label{coor-final-soft}
 \eea
In the paraxial approximation we have $\theta\ll 1$, $|G_2|\ll |G_1|$ and the linear polarization in the scattering plane becomes dominant.

\begin{figure}\center
\includegraphics[scale=0.5] {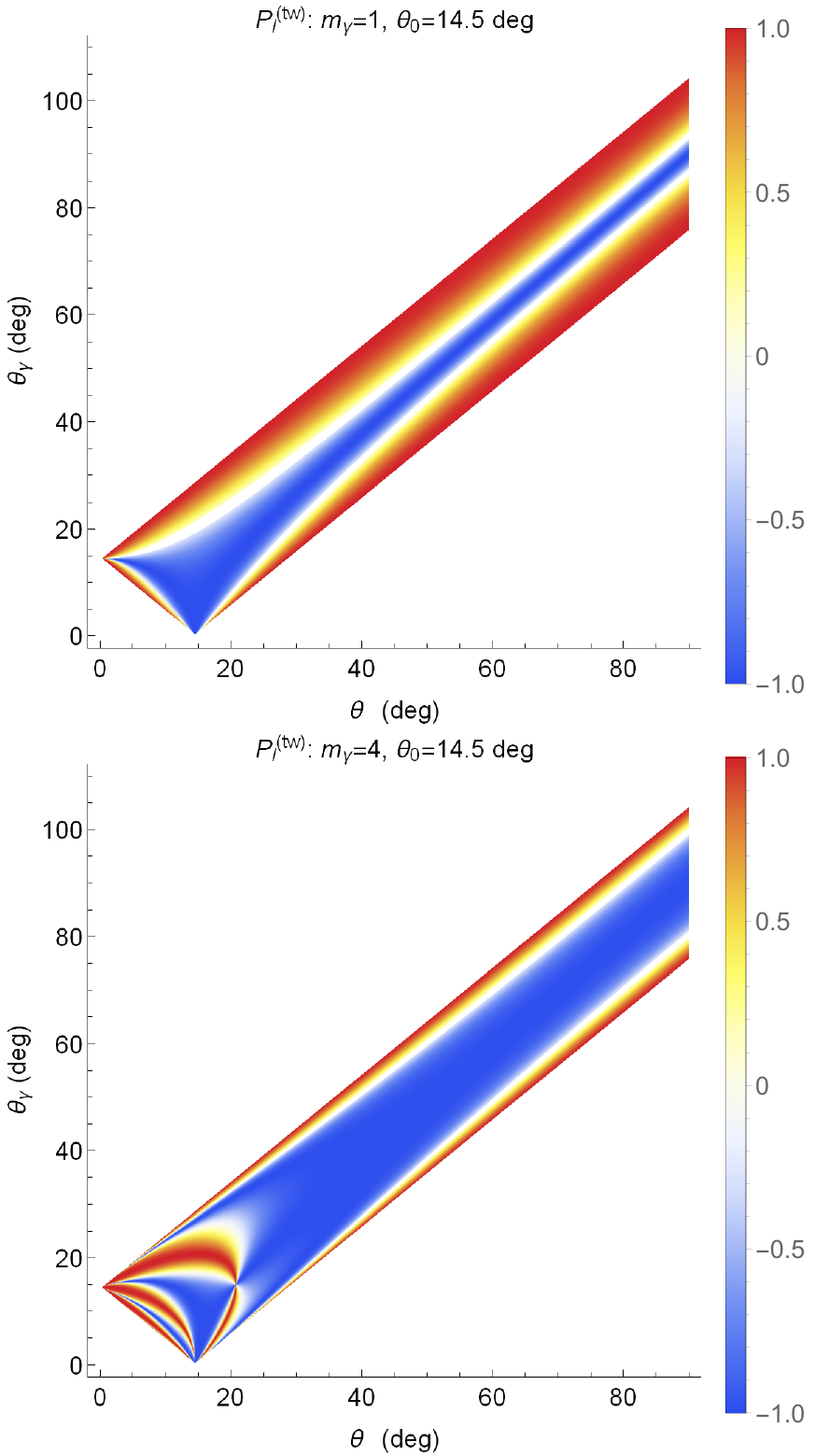}
    \caption{Density plots for the linear polarization $P^{\rm (tw)}_l$, for $\theta_0=14.5 ^{\circ}$ and $m_\gamma = 1$ and $4$.}
    \label{Fig_density_th14}
\end{figure}

{ Similar to the definition of Eq.~\eqref{degreePldef}  we can obtain the degree of linear polarization in the current setup from Eq.~\eqref{coor-final-soft} 
 \be
 P^{\rm (tw)}_l = \fr{|G_1|^2 -|G_2|^2}{|G_1|^2 +|G_2|^2}.
 \label{degreePltw}
 \ee

In Fig.~\ref{Fig_polarization_th14} this expression is plotted as a function of the emitted photon 
angle $\theta_\gamma$ for various values of the vortex electron
opening angle $\theta$ and for $\theta_0 = 14.5^\circ$, which corresponds to $\omega= 2.25$ eV (green light), a refraction index of $n = 1.33$ (water), and the electron kinetic energy of $300$  keV ($v = 0.78$), which is typical for electron microscopes. For completeness, we also show in Fig.~\ref{Fig_density_th14} the
degree of linear polarization as a function of  angles  $\theta$ and  $\theta_\gamma$.
Note, that $P^{\rm (tw)}_l$ always satisfies the condition $-1<P^{\rm (tw)}_l< 1$ and attains its maximal value $P^{\rm (tw)}_l=1$ at the borders of the interval~\eqref{interval} where $\sin\delta=0$. However, its minimal value depends on the relation between $\theta$ and $\theta_0$. For $\theta> \theta_0$, the minimal value $P^{\rm (tw)}_l=-1$ corresponds to $\cos\theta_\gamma = \cos\theta/\cos\theta_0$ and any $m_\gamma$. But when $\theta< \theta_0$, $P^{\rm (tw)}_l$ attains its minimal value $P^{\rm (tw)}_l=-1$ at $\cos\theta_\gamma = \cos\theta_0/\cos\theta$ and only for odd values of $m_\gamma=\pm 1, \pm 3, \pm 5, \dots$, while for zero and even $m_\gamma=0, \pm 2, \pm 4,  \dots$ the same $\theta_\gamma$ results in a maximum, $P^{\rm (tw)}_l=1$.

We see that  $P^{\rm (tw)}_l$ depends essentially on $m_\gamma$ and, therefore, Eq.~\eqref{degreePltw} generalizes the result of Eq.~(36) from Ref.~\cite{ISZ-2016} that has been obtained for the case of the plane-wave photons.}

Finally, we note that at $ \theta \to 0 $ and $m\to \lambda$, the result \eqref{coor-final-soft}  coincides with Eq.~\eqref{Psi-app} up to factor $i^{-\lambda}$ as it should be. To prove that, we must take into account that in the considered limit $\theta_0\pm \theta \approx \theta_0$, therefore, we can replace $\theta_\gamma \to \theta_0 $ everywhere except the function $F(\theta, \theta_\gamma, \theta_0)$ and perform the integration over $\theta_\gamma$ in accordance with Eq.~\eqref{intF}. Using further Eq.~\eqref{limit-psi}, we obtain that
 \bea
&&\psi_{p'_\perp p'_z , m-m_\gamma, \lambda'}(x_e) \vert_{\theta' \to\, 0}  \to \delta_{m-m_\gamma,\lambda} \,  i^{-\lambda}\,   
 U^{(\lambda)}(E, \lambda)  \nn\\
&& \times e^{-i(E t_e - |\bp| z_e)} 
 \eea
 and that
 \be 
C_{ \lambda' \lambda_\gamma m_\gamma} \to  - \delta_{\lambda' \lambda} \sqrt{2\pi \alpha}\, 2vE \lambda_\gamma\, \sin\theta_0.
 \ee 
 As a result, we find
  \be 
\Psi^{\rm (tw)}_{p_\perp p_z m \lambda}(x_e, x_\gamma) \to
i^{-\lambda}\Psi^{\rm (pw)}_{\bp\lambda}(x_e, x_\gamma),
  \ee
where the expression $\Psi^{\rm (pw)}_{\bp\lambda}(x_e, x_\gamma)$ is given by Eq.~\eqref{Psi-app}.

\vspace{5mm}

\section{From VC emission to emission of equivalent photons in Weizs\"acker-Williams method}
\label{Sec_WW}
\begin{figure}\center
\includegraphics[scale=0.3]{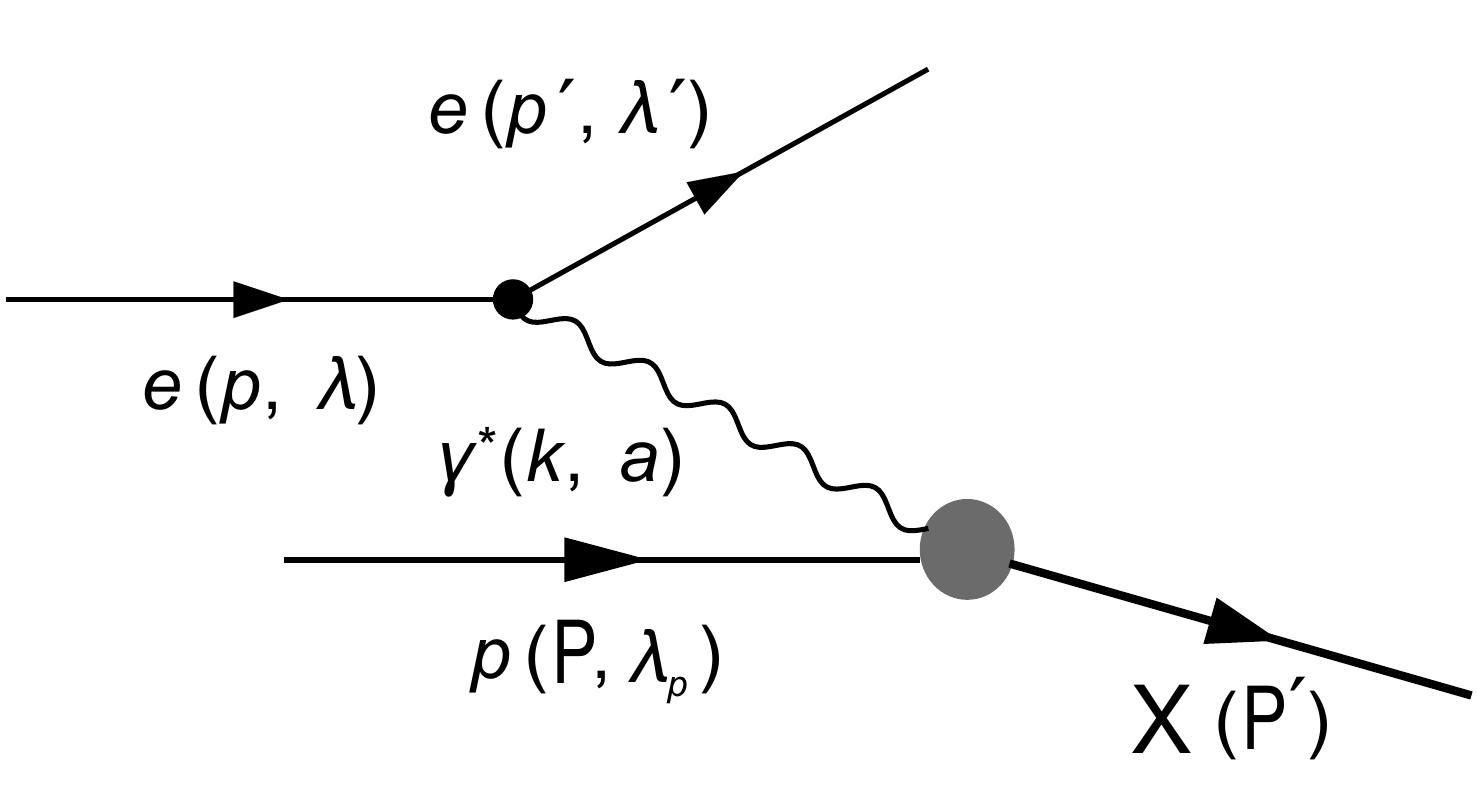}
    \caption{Feynman diagram for the virtual process describing emission and absorption of an equivalent photon of momentum $k$ and polarization $a$.}
    \label{Fig_diag_EP}
\end{figure}

{  The method used in this paper can also be applied to the problem of characterizing the final wave function in the process of radiation of an equivalent photon (EP) by high-energy charged particle. To clarify this point, let us consider the process of inelastic high-energy $ep$-scattering  (Fig.~\ref{Fig_diag_EP})
 \be
 e(p)+p(P)\to e(p')+X(P'),
 \label{DISprocess}
 \ee
where $X$ is the final hadronic state with the total momentum $P'$.

This process may be seen as a two steps reaction:

({\it i}) emission of an EP $\gamma^*$ with momentum $k=p-p'$ and virtuality $k^2< 0$ by the initial electron;

({\it ii}) absorption of this EP by a proton  and  production of the final hadron state $X$.

In the standard approach, the first step of the discussed process corresponds to the virtual photon emission
 \be
 e(p, \lambda)\to e(p', \lambda')+\gamma^*(k, a).
  \label{EP}
  \ee
Here the initial and final electron is described by the plane waves $u_{\bp \lambda} e^{-ipx}$ and
$u_{\bp' \lambda'} e^{-ip'x}$, similar to VC-emission. However, EP is described by the
plane wave $e^{(a)}_\mu e^{-ikx}$ and corresponds to the state with momentum $\bk$, energy $\omega=\sqrt{\bk^2+k^2}$, virtuality $k^2 <0$ and helicity $a =0,\, \pm 1$. Four-vector $e^{(0)}_\mu$ with helicity $a=0$ corresponds to the scalar ($S$) EP, while 4-vectors $e^{(\pm 1)}_\mu$  -- to the transverse ($T$) EP. The latter four-vectors can be chosen in the same form as in VC emission: $e^{(\pm 1)}_\mu= (0,\, {\bf e}_{\bk \lambda_\gamma})$ with
$\lambda_\gamma = \pm 1$.

The amplitude of the process~\eqref{DISprocess} has the form (for details see review~\cite{BGMS-1975})
 \be
 M(ep\to eX) = \sum_{a=0,\pm 1} \fr{J^{(a)}}{-k^2}\, M^{(a)},
 \ee
where $M^{(a)}=M(\gamma^* p\to X)$ is the amplitude for the absorption of this EP by a proton and the electromagnetic part is given by
 \be
 J^{(a)}= (-1)^{a+1} \sqrt{4\pi \alpha}\, \bar u' \gamma^{\mu} u (e^{(a)}_\mu)^*.
 \ee
Clearly, $J^{(\pm)}$ coincides with the amplitude $M_{fi}$ from Eq.~\eqref{Mfist} with coefficients
\bea
 \label{D}
M^{\lambda\lambda' \lambda_\gamma}_{\sigma \sigma_\gamma}&=&-\sqrt{4\pi \alpha}\,
 8\sigma\lambda\delta_{\lambda \lambda'} \sqrt{EE'}
 \, d^{\,1/2}_{\sigma \lambda}(\theta)\,
 \\
 \nn
 &&d^{\,1/2}_{\sigma-\sigma_\gamma, \lambda}(\theta')\,d^{\;1}_{\sigma_\gamma,\,\lambda_\gamma}(\theta_\gamma)\,
   \left(\delta_{0,\, \sigma_\gamma}
 -\sqrt{2}\,\delta_{2\sigma,\sigma_\gamma}\right).
 \eea
taken in the limit $E\gg m_e$ and $E'\gg m_e$.

The discussed cross-section of the inelastic $ep$ scattering with the unpolarized electrons can be presented as (see review~\cite{BGMS-1975})
 \be
 d\sigma_{ep}=dn_T\, \sigma_T + dn_S\, \sigma_S,
  \ee
where $\sigma_T$ ($\sigma_S$) is the cross-section for absorption of a virtual $T$ ($S$) photon by a proton with production of the final hadronic state $X$ the phase volume of which is $d\Gamma_X$,
 \bea
\sigma_T& =& \int  \mbox{$\frac 12$}\left(|M^{(+)}|^2 + |M^{(-)}|^2\right) \fr{d\Gamma_X}{4I},
 \\
\sigma_S& =& \int|M^{(0)}|^2 \fr{d\Gamma_X}{4I},\;\;
  I=\sqrt{(kP)^2-k^2 P^2},
 \nn
 \eea
while  $dn_T$ ($dn_S$) is the number of corresponding virtual photons given by
 \bea
dn_T &=&\mbox{$\frac 12$}\sum_{\lambda\lambda'}\left(\left|J^{(+)}/k^2\right|^2 + \left|J^{(-)}/k^2\right|^2\right) \fr{I}{I_{ep}} \, \fr{d^3 p'}{(2\pi)^3 2E'},
\nn 
 \\
dn_S & =&\mbox{$\frac 12$}\sum_{\lambda\lambda'} \left|J^{(0)}/k^2\right|^2 \fr{I}{I_{ep}} \, \fr{d^3 p'}{(2\pi)^3 2E'},
  \eea
 where $I_{ep}=\sqrt{(pP)^2-p^2 P^2}$.  It is important to note that $dn_T\sim dn_S$ and that $\sigma_S$ disappears  on mass shell.

Below we will distinguish two different cases for this process.

({\it i}) {\it Deep inelastic $ep$ scattering.} In this case the above equations are used in order to extract information about a virtual process $\gamma^* p \to X$. Therefore, we must  take into account the $k^2$-dependence of the cross sections $\sigma_{T}(kP, k^2)$ and $\sigma_{S}(kP, k^2)$. This is a problem we would like to address in a future work.

({\it ii}) {\it Weizs\"acker-Williams (or equivalent photon) approximation.} In the region of small virtuality
of EP, we can neglect the contribution of the scalar EP (since $\sigma_S \to 0$
at $k^2\to 0$) and replace the cross section $\sigma_T$ by its value on the mass-shell
 \be
 \sigma_{T}(kP, k^2)\to \sigma_\gamma(kP),\;{\rm when}\; k^2\to 0.
 \ee
Henceforth, in the equivalent photon approximation (EPA) the cross-section is
 \be
 d\sigma_{ep}=dn_T\, \sigma_\gamma (kP).
 \ee
Certainly, this approximation is valid not only for the inelastic $ep$ scattering, but for a number of processes with different initial and final states. 

As a result, emission of $T$ photons in EPA has a close similarity with the VC radiation.
The most important distinction is related to the differences in kinematics.

In particular, if the initial electron is in the plane-wave state, then its virtuality reads
 \be
 k^2=- \fr{k_\perp^2 +(m_e\omega/E)^2}{1-(\omega/E)}
 \ee
and in the soft photon regime we can use relations
 \be
\omega \ll E, \;\; k^2 \approx - k_\perp^2,\;\; k_z\approx \omega.
 \ee
Now the evolved wave function of the final $e \gamma^*$ state has the form of Eq.~\eqref{Psi-pl-tw} in the coordinate representation.
Therefore, the degree of linear polarization of EP reads
 \be
 P_l^{\rm (pw)}= 1- \fr{\omega^2}{2E^2}.
 \ee
In other words, for EP,  the longitudinal linear polarization
in the plane of the vectors $\bp$ and $\bk$ is dominating in
the region where $\theta' \ll \theta_\gamma$. Moreover, the mean helicity of EP
 \be
 \langle \lambda_\gamma \rangle=\fr{|g_1+g_2|^2-|g_1-g_2|^2}{|g_1+g_2|^2+|g_1-g_2|^2}\approx
 2\lambda\fr{\omega}{E}
 \ee
is proportional to the helicity of the initial electron and is small in the regime considered. Here, the factors $g_1$ and $g_2$ are the ones defined in Eq.~\eqref{pw_g_def}. Both these results are well-known in the ordinary EPA, but their applicability broadens when considered as a result of derivations presented here.

If the initial electron is in the twisted state with TAM projection $m$, then the evolved wave function of the final $e \gamma^*$ state in the coordinate representation has  the form similar to Eq.~\eqref{TW_EV2-1}:
\bea
 \label{TW-EV2-1EP}
 \Psi^{\rm (tw)}_{p_\perp  p_z m\lambda}(x_e, x_\gamma)&=& -{i} \sqrt{2\pi\alpha/E}\,
\sum_{\lambda_\gamma m_\gamma} 
\int d\omega d\theta_\gamma \sin\theta_\gamma
 \nn
\\
&\times&  \sqrt{1-\omega/E}\, F(\theta, \theta_\gamma, \theta_0)
 C_{\lambda_\gamma m_\gamma}
 \\
 &\times& {\bm A}_{k_{\perp}k_z m_{\gamma} \lambda_{\gamma}} (x_{\gamma})\, 
 \psi_{p'_{\perp}p'_z, m-m_\gamma, \lambda} (x_e),
 \nn
 \eea
where
 \bea 
 C_{\lambda_\gamma m_\gamma}&=& 2\lambda \sum_{\sigma \sigma_\gamma} 2\sigma 
 d^{1/2}_{\sigma \lambda}(\theta) d^{1/2}_{\sigma-\sigma_\gamma, \lambda}(\theta')
 \\
 &\times&  d^{\;\;1}_{\sigma_\gamma,\,\lambda_\gamma}(\theta_\gamma)\,
   \left(\delta_{0,\, \sigma_\gamma}
 -\sqrt{2}\,\delta_{2\sigma,\sigma_\gamma}\right) 
 \nn
 \\
  &\times& \cos[(m-\sigma)(\delta-\delta')+(m_\gamma-\sigma_\gamma)\delta']\,.
  \nn 
  \eea
Using this equation, we can easily analyze the polarization of EP following what was just done for the plane-wave initial electron. 

}
\section{Discussion}
\label{Sec_Disc}

{ {\it (i)} First, let us discuss our primary result given by expressions~\eqref{EV2-1} and \eqref{TW_EV2-1}. We have started with consideration of the evolved wave function for the VC process with the plane-wave initial electron.
By construction given in Sect.~\ref{Sec_Evolved_def}, the final wave function is an entangled state that can be regarded as a superposition of plane waves
$\psi_{\bp' \lambda'}(x_e)$ and ${\bf A}_{\bk \lambda_\gamma}(x_\gamma)$ of two final particles:
  \be
  \Psi_{\bp \lambda}(x_e, x_\gamma) =\int \sum_{\lambda' \lambda_\gamma} B_{\lambda' \lambda_\gamma}\,
  \psi_{\bp' \lambda'}(x_e) {\bf A}_{\bk \lambda_\gamma}(x_\gamma)\, d\Gamma,
  \ee
where the coefficients $B_{\lambda' \lambda_\gamma}$ are proportional to the  plane-wave 
matrix element of the process and $d\Gamma$ is the phase  volume of the final particles. 
In the standard interpretation, the energy-momentum conservation law enclosed in this expression plays a crucial role. If during the detection of, say, the final electron, we measure its momentum, $p'$, then the photon automatically turns out to be a plane-wave state with momentum $k=p-p'$. Note, that the detector here needs not to be fully specified, it only serves as an instrument of projection on the plane-wave state of the final electron.

However, if we expand the plane waves of final particles into the twisted (Bessel) waves moving along the $z$-axis and perform a trivial integration over the azimuthal angle,
we obtain the final two-particle wave function of Eq.~\eqref{EV2-1}. Now, it appears as a superposition of twisted waves, the TAM projections of which onto the $z$-axis, $m'$ and $m_\gamma$, run through an infinite series of values. If the detector projects this superposition onto a twisted wave electron state with a definite projection $m'$, then only one term in the sum \eqref{EV2-1} survives. The final photon automatically turns out to be in the state that is a twisted wave with a definite projection
  \be
m_\gamma=\lambda-m',
  \ee
given by the TAM conservation law. As above, the detector design is of little importance, only its ability to project onto a twisted state matters. 
In particular, in a generalized measurement scheme this may be a detector that makes measurements with large uncertainties for the azimuthal angle of the final electron, as discussed in papers~\cite{KBGSS-2022, KBGSS-2023}. 

These considerations are easily generalized for the process in which the initial state is not a plane wave with the helicity $\lambda$, but a Bessel wave with the definite TAM projection $m$ 
(see Eq.~\eqref{TW_EV2-1}). In this case, the conservation law of TAM requires
  \be
m_\gamma=m-m',
  \ee
if the detector projects the final electron from the two-particle state onto a twisted wave with a definite projection $m'$.
Analogously, for the detector measuring a definite $m_\gamma$, we reconstruct the final electron as a twisted state with 
  \be
m'= m-m_\gamma.
  \ee

{\it (ii)} Let us now turn to the properties of EP which is produced by some plane-wave initial electron with helicity $\lambda$, and follow the lines of the analysis above. 
If the detector projects the $e\gamma^*$ state onto the final electron with a definite projection $m'$, i.e. a twisted wave, then the final EP will also be in the twisted state with a definite projection $m_\gamma=\lambda-m'$. Further, this EP moves along the $z$-axis and experiences a head-on collision with a plane-wave proton having helicity $\lambda_p$. As a result, the
final hadron state $X$ will be in the twisted state with the TAM projection 
 \be
 m_X=m_\gamma-\lambda_p 
 \ee
and its properties could be quite different from the ordinary final hadron state resulting from the $ep$-scattering.

Should we consider a twisted initial electron
with the definite TAM projection $m$ and use the detector that projects the final two-particle $e\gamma^*$ state onto a twisted wave electron state with a definite projection $m'$, then the TAM projection of EP will be $m_\gamma=m-m'$ as is illustrated by Eq.~\eqref{TW-EV2-1EP}.

}

{\it (iii)} The question in what processes twisted photons could be emitted is both theoretically and experimentally challenging. 
Let us briefly recall the known basic methods for the production of twisted photons. In paper~\cite{Sasaki-McNulty-2008} it was shown that when an electron is moving in a helical undulator, it emits the photons of the second harmonic in the twisted state. A little later it was shown in paper~\cite{Afanasev-Michailichenko-2009} that higher harmonics have the same property. Then in 2011, it was shown in papers~\cite{jentschura-serbo-2011-1}, \cite{jentschura-serbo-2011-2} and \cite{ivanov-serbo-2011} that in the Compton back scattering of twisted laser photons by relativistic electrons, the final high-energy photon is also twisted in the main region of the emitted angles. The twisted photons with the energy 99 eV have been produced by utilizing a helical undulator at the synchrotron light source BESSY II \cite{bahrdt2013}.
In 2015--2017, the experiments carried out at BNL on the nonlinear Compton effect~\cite{Brookhaven-experiment-2015} were interpreted in~\cite{Taira-I-2017} as an  observation of the second harmonic which is twisted. In this case, the circularly polarized laser light plays role of the spiral undulator in the {BESSY~II} experiments. A more general study~\cite{Taira-II-2017} shows that twisted photons are emitted at the harmonics higher than the first one during the spiral motion of an electron.

In this respect, the VC emission, specifically in the case of an initial plane-wave electron, can be regarded as a new technique for generating the twisted photons. Recently, it was argued in Refs.~\cite{KBGSS-2022, KBGSS-2023} that the use of generalized measurements is indeed an effective method for obtaining twisted photons. Our results follow up these works and give an interesting perspective for non-conventional measurement schemes.

{  {\it (iv)} Finally, 
as means to provide an unburdened version of the mathematics involved while considering the process of VC radiation, we also refer the reader to Appendix~\ref{App_C}.
}

\section{Conclusions}
In this paper we explored the application of the evolved state formalism in the processes of VC emission by a plane wave and Bessel wave (i.e., carrying OAM) electron.
This way we obtained the quantum mechanically complete description of the final entangled state of the photon and the electron independent of the detection scheme or the properties of a detector. We emphasize that the final system is an entangled state of two particles of different type, however, the obtained form of the wave-function is such that both particles contribute symmetrically.

Besides, we considered the limiting cases of low-energy radiation and high-energy initial electron confirming that the evolved state reproduces correctly the expected behaviour.
We also verify the results of paper~\cite{ISZ-2016} that the photonic part of the final state for the case of the twisted initial electron has the angular distribution and polarization which  differ considerable from the ordinary VC radiation. 

{ In addition, we analyze the wave function of the final state in the process of equivalent photon emission in the Weizs\"acker-Williams
approximation. }

\begin{acknowledgments}
We are thankful to I. Ivanov for useful discussions.
The studies in Sec. II are supported
by the Russian Science Foundation (Project No. 21-42-04412; https://rscf.ru/en/project/21-42-04412/). The studies in Sec. III are supported by the Government of the Russian Federation through the ITMO Fellowship and Professorship Program. The studies in Sec.IV are supported by the Russian Science Foundation (Project No.\,23-62-10026;  https://rscf.ru/en/project/23-62-10026/).
{ The studies in Sec.V are supported by the Ministry
of Science and Higher Education of the Russian Federation (agreement No. 075-15-2021-1349).}
\end{acknowledgments}

\appendix

\section{Properties of twisted electrons and photons }
\label{App_tw_properties}
In this Appendix we collect some useful formulae related to properties of twisted electrons and photons (for more
detail see reviews~\cite{Blioch-Ivanov-2017} and~\cite{Knyazev-Serbo-2018}, respectively).

\subsection*{Electrons}

The bispinor $u_{\bp' \lambda'}$ has the following evident
dependence on the spherical angles $\theta'$ and $\varphi'$ (see Eq. (A1) from~\cite{ISZ-2016}):
 \be
u_{\bp' \lambda'}=\sum_{\sigma'=\pm 1/2}
 d^{1/2}_{\sigma' \lambda'}(\theta')
  \,U^{(\sigma')}(E',\lambda')\,e^{-i\sigma' \varphi'},
  \label{function u}
 \ee
where the basis bispinors $U^{(\sigma')}(E',\lambda')$ are expressed as follows:
 \bea
 \label{eq_U_spinor}
 U^{(\sigma')}(E', \lambda') = \left( \begin{array}{c}
                                   \sqrt{E' + m_e} \, w^{(\sigma')} \\[2mm]
                                   2 \lambda' \sqrt{E' - m_e} \, w^{(\sigma')}
                                   \end{array}
                          \right),\\ \nn
   w^{(+1/2)}= \left(
   \begin{tabular}{c}
     1  \\
     0  \\
     \end{tabular}
   \right),\;\;
   w^{(-1/2)}= \left(
   \begin{tabular}{c}
     0  \\
     1  \\
     \end{tabular}
   \right)\,.
   \eea
They do not depend on the direction of $\bp^\prime$ and are eigenstates of the spin projection operator $\hat s'_z$ with eigenvalues $\sigma' = \pm 1/2$. Note also that the expression $u_{\bp' \lambda'}\, e^{im' \varphi'}$ is the eigenfunction 
in the momentum representation of the TAM operator  $\hat j'_z$ with the eigenvalue $m'$ :
 \be
\hat j'_z\, u_{\bp' \lambda'}\, e^{im' \varphi'} =
  m' \,u_{\bp' \lambda'}\, e^{im' \varphi'}.
 \label{e-eigen}
 \ee

The function
 \bea 
 \label{bessel-wave-electron1}
 \psi_{p'_\perp p'_z  m'\lambda'}(x) &=&
 \int_0^{2\pi} i^{-m'}\, u_{\bp' \lambda'}\, e^{im'\varphi'} \, e^{-i p' x}\fr{d\varphi'}{2\pi}
   \\
   \nn
  &=&e^{-i(E' t - p'_z z)} \sum_{\sigma'=\pm 1/2}
 i^{-\sigma'} d^{\;1/2}_{\sigma'\,\lambda'}(\theta')\\ \nn
 &\times&
 J_{m'-\sigma'}(p'_\perp r_\perp)\,e^{i(m'-\sigma')\varphi_r}\,
 U^{(\sigma')}(E', \lambda')
     \eea
 corresponds to the twisted electron with longitudinal momentum $p'_z$, transverse momentum modulus $p'_\perp$, energy $E'=\sqrt{p'^2_\perp+p'^2_z+m^2_e}$,
projection of the electron TAM onto the $z$-axis equal  $m'$ and helicity $\lambda'=\pm 1/2$. In the  paraxial approximation the above sum is dominated by the term with $\sigma'=\lambda'$:
\bea \nn
 \psi_{p'_\perp p'_z m' \lambda'}(x) \approx
   i^{-\lambda'}  \cos(\theta'/2)\,
 J_{m'-\lambda'}(p'_\perp r_\perp)\\
 e^{i(m'-\lambda')\varphi_r}\,
 U^{(\lambda')}(E', \lambda')\,e^{-i(E' t - p'_z z)}.
   \label{paraxial-bessel-wave-electron}
   \eea
 Finally, in the limit  $\theta' \to 0$ , we obtain
\be
 \psi_{p'_\perp p'_z m'\lambda'}(x) \vert_{\theta' \to\, 0} \to \delta_{m'\lambda'} \,  i^{-\lambda'}\,   
 U^{(\lambda')}(E', \lambda')\, e^{-i(E' t - p'_z z)}  \,.
  \label{limit-psi}
      \ee
 i.\,e. in this limit and for $ m' = \lambda'$, the wave function of a twisted electron coincides with a plane wave along the $ z $ axis up to the phase factor $ i^{-\lambda'} $.

{ Similarly, the function  
\be 
 \label{bessel-wave-electron2}
 \psi_{p_\perp p_z  m\lambda}(x) =
 \int_0^{2\pi} i^{-m}\, u_{\bp \lambda}\, e^{im\varphi} \, e^{-i p x}\fr{d\varphi}{2\pi}
 \ee
corresponds to the initial twisted electron with longitudinal momentum $p_z$, transverse momentum modulus $p_\perp$, energy $E=\sqrt{p^2_\perp+p^2_z+m^2_e}$,
projection of the electron TAM onto the $z$-axis equal  $m$ and helicity $\lambda=\pm 1/2$.  In the limit $\bp_\perp=\theta=0$, momentum $\bp=(0, 0, |\bp|)$,  the $z$-projection of the orbital angular momentum disappears because in this limit $\hat l_z\propto \hat {\bf l} \cdot \bp=0$,  and the state helicity coincides with the $z$-projection of TAM. Indeed, if    $\theta \to 0$, we have
\be
 \psi_{p_\perp p_z m\lambda}(x) \vert_{\theta \to\, 0} \to \delta_{m\lambda} \,  i^{-\lambda}\,   
 U^{(\lambda)}(E, \lambda)\, e^{-i(E t - p_z z)}  \,.
  \label{limit-psi2}
      \ee
i.\,e. in this limit  the wave function of  the initial electron coincides (up to the phase factor $ i^{-\lambda} $) with a plane wave along the $ z $ axis. Moreover, this plane wave has the same eigenvalue $\lambda$ both for  $\hat j_z$ and helicity operators, since in this limit $\hat j_z=\hat s_z$. 
 }

\subsection*{Photons}

The evident dependence of the photon polarization vector on spherical angles $\theta_\gamma$ and $\varphi_\gamma$ reads (see detail in Ref.~\cite{MSSF-2013}):
 \be
 \bee_{\bk \lambda_\gamma}=
 \sum_{\sigma_\gamma=0,\pm 1}
 d^{\;\;1}_{\sigma_\gamma \lambda_\gamma}(\theta_\gamma)
  \,\bm \chi_{\sigma_\gamma}\,e^{-i\sigma_\gamma \varphi_\gamma},
  \label{function e}
 \ee
where the basis vectors
  \be
   {\bm \chi}_{0}= \left(0,\, 0,\, 1 \right),\;\;
   {\bm \chi}_{\pm 1}= \mp \fr{1}{\sqrt{2}} \left(1,\,\pm i,\, 0   \right)
\label{chi}
 \ee
represent the eigenstates of the photon spin $z$-projection operator $\hat s^\gamma_z$ with the eigenvalues
$\sigma_\gamma=0,\,\pm 1$.

The function
 \bea \displaystyle
{\bf A}_{k_\perp k_z  m  \lambda_\gamma}(x) &=&
 \int_0^{2\pi} i^{-m}\, {\bf e}_{\bk \lambda_\gamma}\, e^{im\varphi_\gamma} \, e^{-i k x}\fr{d\varphi_\gamma}{2\pi}
 \label{bessel-wave-evident-0}
 \\ \nn
  &=&e^{-i(\omega t - k_z z)} \sum_{\sigma_\gamma=0;\pm 1}
 i^{-\sigma_\gamma} d^{\;1}_{\sigma_\gamma\,\lambda_\gamma}(\theta_\gamma)\\
 &\times&
 J_{m-\sigma_\gamma}(k_\perp r_\perp)\,e^{i(m-\sigma_\gamma)\varphi_r}\,
 \bm \chi_{\sigma_\gamma},
  \label{bessel-wave-evident}
 \eea
where $J_n(x)$ is the Bessel function of the first kind, corresponds to the twisted photon with longitudinal momentum $k_z$, transverse momentum modulus $k_\perp$, energy $\omega=\sqrt{k^2_\perp+k^2_z}\,/n$,
projection of the photon TAM onto the $z$-axis equals to the integer number  $m$ and helicity $\lambda_\gamma=\pm 1$ \cite{Knyazev-Serbo-2018}. This function is normalized by the condition
 \bea \nn
 \int {\bf A}^*_{k'_\perp k'_z m' \lambda'_\gamma}(x)\,
 {\bf A}_{k_\perp  k_z m \lambda_\gamma}(x)\,d^3r
 \\=\fr{4\pi^2}{k_\perp}\,\delta(k'_\perp-k_\perp)
 \delta_{m' m}\,\delta(k'_z-k_z)\,\delta_{\lambda'_\gamma\lambda_\gamma}.
 \eea
 Note also that the expression ${\bf e}_{\bk \lambda_\gamma}\, e^{im_\gamma \varphi_\gamma}$ is the eigenfunction 
in the momentum representation of the TAM operator  $\hat j^\gamma_z$ with the eigenvalue $m_\gamma$ :
 \be
\hat j^\gamma_z\, {\bf e}_{\bk \lambda_\gamma}\, e^{im_\gamma \varphi_\gamma} =
  m_\gamma \,{\bf e}_{\bk \lambda_\gamma}\, e^{im_\gamma \varphi_\gamma}
 \label{mr-eigen}
 \ee

For small values of the conical angle $ \theta_\gamma $ (which corresponds to the so-called
{\it paraxial approximation})
the sum~\eqref{bessel-wave-evident} is dominated by the term with $\sigma_\gamma=\lambda_\gamma$:
 \bea \nn
 {\bf A}_{k_\perp  k_z m \lambda_\gamma}(x) \approx
 i^{-\lambda_\gamma}\,\cos^2(\theta_\gamma/2)\,
 J_{m-\lambda_\gamma}(k_\perp r_\perp)\\
 \times\,\bee^{i(m-\lambda_\gamma)\varphi_r}\,
 \bm \chi_{\lambda_\gamma}\, \bee^{-i(\omega t-  k_z z)}.
 \label{paraxial-bessel-wave-evident}
 \eea
Therefore in this approximation, the $ z $-projection $m$ of the photon TAM is unambiguously made up of the spin angular momentum projection, approximately equal to $ \lambda_\gamma $, and the projection of the OAM, approximately equal to $ m- \lambda_\gamma $.  Finally, in the limit $ \theta_\gamma \to 0 $ (in this case, $ k_\perp \to 0 $, $ k_z \to k = \omega n $, $ d^{\; 1}_{\sigma \lambda_\gamma} (\theta_\gamma) \to \, \delta_{\sigma \lambda_\gamma} $ and
$ J_ {m- \sigma} (k_\perp r_\perp) \to \delta_{m \, \sigma} $) we obtain
 \be
 {\bf A}_{k_\perp  k_z m \lambda_\gamma}(x)\vert_{\theta_\gamma \to\, 0} \to \delta_{m\lambda_\gamma} \,
 i^{-\lambda_\gamma}\,\bm \chi_{\lambda_\gamma} \,
 \bee^{-i(\omega t-  k_z z)},
 \ee
i.\,e. in this limit and for $ m = \lambda_\gamma $, the wave function of a twisted photon coincides with a plane wave along the $ z $ axis up to the phase factor $ i^{-\lambda_\gamma} $.

In the momentum representation, the wave function of a twisted photon has a particularly simple form

\bea
&&\tilde {\bf A}_{k_\perp  k_z m \lambda_\gamma}({\bf K})=\int {\bf A}_{k_\perp k_z m  \lambda_\gamma}(x)\,
 e^{-i{\bf K} \br} d^3r
 \\
  &&=\fr{4\pi^2}{k_\perp} \delta(K_\perp-k_\perp)\,
 \delta(K_z-k_z)\,i^{-m} {\bf e}_{{\bf K} \lambda_\gamma}\,   e^{-im\varphi_\gamma-i\omega t}
 \nn
  \eea
and it corresponds to plane waves concentrated on the cone with the polar angle $\theta_\gamma=\arctan(k_\perp/k_z)$.

\subsection*{Photon polarizations}
To analyze the polarization of the photon, it is also convenient to introduce the following linear combinations of vectors $ {\bf e}_{\bk \lambda_\gamma} $ with helicities $\lambda_\gamma = \pm 1$: 
 \bea
 \label{paral}
 {\bf e}_{\parallel}&=& \fr{-1}{\sqrt{2}}\left({\bf e}_{\bk, 1}-{\bf e}_{\bk, - 1} \right)\\
 \nn &=&
  \left(\cos\theta_\gamma \cos\varphi_\gamma,\, \cos\theta_\gamma \sin\varphi_\gamma,\,-\sin\theta_\gamma\right),
  \\ \label{perp}
 {\bf e}_{\perp}&=& \fr{i}{\sqrt{2}}\left({\bf e}_{\bk, 1}+{\bf e}_{\bk, - 1} \right)\\
 \nn &=&  \left(-\sin\varphi_\gamma,\, \cos\varphi_\gamma,\,0\right).
    \eea
It is easy to verify that these vectors are mutually orthogonal and orthogonal to the vector $ \bk $. The vector
$ {\bf e}_ \parallel $ defines the longitudinal linear polarization (lying in the scattering plane given by the $ z $-axis and the vector $ \bk $) and the vector $ {\bf e}_{\perp} $  -- the  linear polarization which is orthogonal to the scattering plane. 
A twisted photon state with either a longitudinal $ l = \parallel $ or an orthogonal $ l = \perp $ polarization looks similar to Eq.~\eqref{bessel-wave-evident-0}:
   \be
 {\bf A}_{k_\perp k_z m l}(x) = \int_0^{2\pi}  i^{-m} {\bf e}_{ l}\, e^{i m \varphi_\gamma}\,e^{- i k x}\fr{d\varphi_\gamma}{2\pi}.
 \label{bessel-wave-l}
 \ee

The following relations are exploited in the paper:
 \be 
\sum_{\lambda_\gamma} \lambda_\gamma {\bf e}_{\bk \lambda_\gamma}=-\sqrt{2}\,{\bf e}_{\parallel},\;\;
\sum_{\lambda_\gamma}  {\bf e}_{\bk \lambda_\gamma}=- i\sqrt{2}\,{\bf e}_{\perp},
 \label{e-to-e}
\ee
and could be checked straightforwardly.
\section{Summation over $\lambda'$ and $\lambda_\gamma$ in Eq.~\eqref{phi-st}} 
\label{App_helicity_sum}
Here we calculate the sum
 \be 
{\bf S}= \sum_{\lambda' \lambda_\gamma} u_{\bp' \lambda'} 
{\bf e}_{\bk \lambda_\gamma} \left (M^{\lambda\lambda' \lambda_\gamma}_{\lambda, 0} - 
  M^{\lambda\lambda' \lambda_\gamma}_{\lambda, 2\lambda}\right ),
 \ee 
the expression in parentheses as in Eq.~\eqref{Mat}. First, using Eq.~\eqref{e-to-e}, we find the partial sums
\bea 
\sum_{\lambda_\gamma} {\bf e}_{\bk \lambda_\gamma} d^{\;(1)}_{0 \lambda_\gamma}(\theta_\gamma)&=& - {\bf e}_\parallel \sin\theta_\gamma,\\
\sum_{\lambda_\gamma} {\bf e}_{\bk \lambda_\gamma} d^{\;(1)}_{2\lambda, \lambda_\gamma}(\theta_\gamma)&=& 
\fr{-1}{\sqrt{2}}\left( {\bf e}_\parallel 2\lambda \cos\theta_\gamma-i {\bf e}_\perp \right) .
\eea
Using these expression we obtain
\bea \nn
{\bf S}&=&\sqrt{4\pi \alpha} E_{\lambda \lambda} u_{\bp' \lambda}
\left[{\bf e}_\parallel \sin(\theta_\gamma+
\mbox{$\frac 12$}\,\theta') +i {\bf e}_\perp  2\lambda
\sin( \mbox{$\frac 12$}\,\theta') 
\right]
\\
&+&\sqrt{4\pi \alpha} E_{\lambda,- \lambda} u_{\bp' ,-\lambda}
\bigg[{\bf e}_\parallel 2\lambda \cos(\theta_\gamma+
\mbox{$\frac 12$}\,\theta') \\
&+&i {\bf e}_\perp  \cos( \mbox{$\frac 12$}\,\theta') 
\bigg], \nn
\eea 
with $E_{\lambda \lambda'}$ defined in Eq.~\eqref{Ell}.
In the limiting cases we have
\be
{\bf S}=4\sqrt{\pi \alpha} vE  u_{\bp' \lambda}
{\bf e}_\parallel \sin\theta_\gamma
\ee
for the soft-photon approximation  and
\be
{\bf S}=4\sqrt{\pi \alpha E E' }  u_{\bp' \lambda}
\left[{\bf e}_\parallel \sin(\theta_\gamma+
\mbox{$\frac 12$}\,\theta') +i {\bf e}_\perp  2\lambda
\sin( \mbox{$\frac 12$}\,\theta') 
\right]
\ee
for the ultra-relativistic electron approximation.


\section{Three scalar fields}
\label{App_C}
In this appendix we would like to use the developed formalism and in a simplified toy model of $1 \rightarrow 2$ process with three scalar particles.
For definiteness, consider a process of a scalar particle with mass $M$ and momentum $p$ decaying into two scalar particles with smaller masses $\mu'$ and $\mu''$ ($\mu'+\mu''\leq M$) and momenta $p'$ and $p''$. The tree level matrix element is given by
\be
S^{(1)}_{fi} = -i \lambda (2\pi)^4\,N\, \delta(p'+p''-p),\ 
\ee
$N=\frac1{\sqrt{2E}\sqrt{2E'}\sqrt{2E''}}$, and should be inserted into the definition of the evolved wave function in coordinate representation
\bea \label{app_coord}
&& \la 0| \hat{\phi}(x'') \hat{\phi}(x')|\phi,\phi\ra^{(\text{ev})}   
 \\\
 && =\int\frac{d^3p'}{(2\pi)^3}\frac{d^3p''}{(2\pi)^3}\,
\fr{1}{\sqrt{2 E'}}\,e^{-ip'x'}\,
\fr{1}{\sqrt{2 E''}}\, e^{-i p''x''}\;
 S_{fi}^{(1)}.
 \nn
 \eea
Using delta function to eliminate the integral over $p''$ leads to reduction
\bea 
\int d\Gamma &=&  (2\pi)^4 \int \delta(p'+p''-p)\,\fr{d^3 p'}{(2\pi)^3}\fr{d^3 p''}{(2\pi)^3}\\
&\rightarrow&\ \int\delta(E-E'-E'') |p'|^2 \sin \theta' \frac{d |p'| d\theta'd\varphi'}{(2\pi)^2}, \nn
\eea
and writing down the momenta conservation law $\bp''=\bp-\bp'$ transforms the energy conservation $\delta(E-E'-E'')$ into
$$\delta(E-E'-E'')=\frac{E''}{|p'||p|}\delta(\cos\theta_{pp'}-\cos\theta_0),$$
where $\cos\theta_0=\frac 1{vv'}\left(1-\frac{M^2+\mu'^2-\mu''^2}{2EE'}\right)$, $|p|=vE$, $|p'|=v'E'$. In case the initial particle moves along the $z$ direction, $\cos\theta_{pp'}=\cos \theta'$ and the integral over $\theta'$ can be lifted imposing henceforth $\theta'=\theta_0$.

After substituting series expansions in terms of cylindrical waves 
\bea
&e^{-i p' x'} = \sum\limits_{m' = -\infty}^{+\infty} i^{m'}\, e^{-im'\varphi'} \phi_{p'_{\perp}p'_z m' } (x'),\\
&e^{-i p''x''} = \sum\limits_{m'' = -\infty}^{+\infty} i^{m''}\, \  e^{-im''\varphi''} \phi_{p''_{\perp}p''_z m'' } (x''),\\
&\phi_{p_{\perp}p_z m} (x)=e^{-iEt} J_m(p_{\perp} r_\perp)e^{i(m\phi_r+k_z z)}
\eea
into the evolved wave function, and considering that momenta conservation allows only for $\varphi''=\varphi'\pm \pi$ we can readily evaluate the remaining azimuthal integral in Eq.~(\ref{app_coord}).
Finally, we obtain a simple expression
\bea
&& \la 0| \hat{\phi}(x'') \hat{\phi}(x')|\phi,\phi\ra^{(\text{ev})}  =\frac{i \lambda}{4\pi v} \frac1{(2E)^{3/2}}
 \\\
 &&  \times
 \int \frac{p' dp'}{E'}\, 
\sum_m(-1)^m  \phi_{p'_{\perp}p'_z m } (x')\phi_{p'_{\perp},p_z-p'_z, -m } (x'').
 \nn
 \eea
This could be generalized for the case of initial twisted particle with OAM $m$ and manifest the conservation law $m= m'+m''$.
  
\bibliographystyle{unsrt}

\end{document}